\documentclass[amssymb,useAMS,prd,aps,amsmath,superscriptaddress,nofootinbib,twocolumn,floatfix]{revtex4}
\usepackage{color}
\usepackage{axodraw4j}
\usepackage{pslatex}
\usepackage{pdfpages}
\usepackage{graphicx}
\usepackage{psfrag}
\usepackage{pdftricks}
\usepackage{multirow}
\usepackage{graphicx}
\usepackage{hyperref}
\begin{document}
\def\mlsp{m_{\chi_1^0}}
\def\ra{\rightarrow}
\title{The 125 GeV Higgs in the NMSSM in light of LHC results and astrophysics constraints}

\author{Daniel Albornoz V\'asquez}
\affiliation{Institut d'Astrophysique de Paris, UMR 7095 CNRS, Universit\'e Pierre et
Marie Curie, 98 bis Boulevard Arago, Paris 75014, France}

\author{Genevi\`eve B\'elanger}
\affiliation{LAPTH, U. de Savoie, CNRS,  BP 110,
  74941 Annecy-Le-Vieux, France.}

\author{C\'eline B\oe hm}
\affiliation{Institute for Particle Physics Phenomenology, University of Durham, Durham, 
DH1 3LE, UK}
\affiliation{LAPTH, U. de Savoie, CNRS,  BP 110,
  74941 Annecy-Le-Vieux, France.}

\author{Jonathan Da Silva}
\affiliation{LAPTH, U. de Savoie, CNRS,  BP 110,
  74941 Annecy-Le-Vieux, France.}
\affiliation{Institute for Particle Physics Phenomenology, University of Durham, Durham, 
DH1 3LE, UK}
\author{Peter Richardson}
\affiliation{Institute for Particle Physics Phenomenology, University of Durham, Durham, 
DH1 3LE, UK}

\author{Chris Wymant}
\affiliation{Institute for Particle Physics Phenomenology, University of Durham, Durham, 
DH1 3LE, UK}

\begin{abstract}
Recent LHC data suggest an excess in the Higgs decay channels into $\gamma \gamma$ and $Z Z$ at $\sim$125~GeV. 
The current excess in the diphoton channel is twice that expected from a Standard Model Higgs; whilst this may well change with more statistics, it is interesting to consider the implications should the result persist.
Here, we assess whether the NMSSM with a neutralino dark matter candidate could explain this excess when astrophysical constraints (e.g.  no overproduction of gamma rays and radio emission in the galaxy, no anomalous excess in the dark matter direct detection experiments and no dark matter overabundance) are imposed on the neutralino. This enables us to disregard unphysical regions of the parameter space even though the Higgs signal is compatible with the observed excess. The result of our analysis is that there are configurations of the parameter space which can explain the signal strength reported by the ATLAS and CMS collaborations for a Higgs mass within the required range. 
Should the observed signal strength finally be compatible with Standard Model expectations, it would be difficult to distinguish between the discovery of Standard Model Higgs and a SM-like Higgs from the NMSSM, unless one performs dedicated searches of very light Higgs bosons and possibly investigate peculiar signatures of supersymmetric particles. We also propose a new jets + missing $E_T$ signal for the case where the LSP is a singlino-like neutralino.
\end{abstract}

\maketitle

\section{Introduction}
\label{sec:intro}

The LHC collaborations have reported new results on the search for the standard model Higgs, $h$ \cite{ATLAS:2012ae,Chatrchyan:2012tx}. By combining analyses, the ATLAS collaboration has excluded the SM Higgs 
 in the mass range $112.7~{\rm GeV}<M_h<115.5~{\rm GeV}$ and $131~{\rm GeV}<M_h<453~{\rm GeV}$  at 95\%CL 
with an integrated luminosity up to $4.9 fb^{-1}$.
Similarly, the CMS collaboration excluded a SM Higgs in the range
$127~{\rm GeV}<M_h<600~{\rm GeV}$  at 95\%CL with integrated luminosity of $4.6-4.7 fb^{-1}$.
Furthermore, ATLAS reported an excess at 2.8$\sigma$ in the $h\rightarrow \gamma\gamma$  decay channel, consistent with a Higgs mass $M_h=126$~GeV \cite{ATLAS:2012ad}. This signal is larger than expected from a SM Higgs (with signal strength $\sigma/\sigma_{SM}=2\pm 0.8$) and might favour new physics. 
An excess  in the $\gamma \gamma$ channel was also reported by the CMS experiment \cite{Chatrchyan:2012tw} but at a slightly lower mass $M_h=124$~GeV.

Although ATLAS and CMS statistics are not sufficient to claim a Higgs  discovery, it is interesting to assume  that these are possible evidence for the Higgs ~\cite{Baer:2011ab,Heinemeyer:2011aa,Carena:2011aa,Arbey:2011ab,Olive:2012it,Espinosa:2012ir,Carmi:2012yp,Azatov:2012bz,Baer:2012uy,King:2012is,Gunion:2012zd,Brooijmans:2012yi,Cao:2011sn,Christensen:2012ei,Desai:2012qy} and investigate whether this is compatible with the predictions which arise in supersymmetric extensions of the Standard Model such as the Minimal Supersymmetric Standard Model (MSSM) or Next-to-MSSM (NMSSM). To fit the data, the Higgs candidate must have both a mass in the favoured range (that is,  
[122,128]~GeV when taking into account the theoretical and experimental uncertainties~\cite{Heinemeyer:2011aa})  and a signal strength that is consistent with the value of $\sigma/\sigma_{SM}$   measured by ATLAS and CMS.

In the MSSM, the  Higgs mass is required to be light $M_h \leq 135{\rm GeV}$~\cite{Djouadi:2005gj}. While a Higgs mass of 125 GeV or larger  is still possible it requires  
appreciable fine-tuning~\cite{Hall:2011aa}. In addition, the question of whether the signal strength can be as large as suggested by the data must be answered.  Assuming that the Higgs is produced by gluon fusion, the signal strength is defined by the ratio 
$$R_{ggXX} =  \frac{\sigma(gg\rightarrow h)_{BSM} BR(h\rightarrow X X)_{BSM}}{\sigma(gg\rightarrow h)_{SM} BR(h\rightarrow X X)_{SM}}$$ 
where $X$ can be either photons, $Z$ or $W$. It is generally expected that, after applying other constraints on the model,  $R_{gg \gamma \gamma} $ is at most as large as unity~\cite{Arbey:2011aa,Brooijmans:2012yi,AlbornozVasquez:2011aa}. Indeed, the large mixing in the stop sector (which is required to have a large enough Higgs mass to fit the data) means that the stop contribution to the Higgs-gluon-gluon ($hgg$) loop-induced vertex interferes destructively with the dominant top contribution. This reduces the Higgs production cross section and is not completely compensated by the increase in the Higgs-$\gamma$-$\gamma$ coupling (even though an exception was found in a corner of the parameter space where heavily mixed staus can lead to an increase in the  $h \rightarrow \gamma\gamma$ partial decay width ~\cite{Carena:2011aa}). Large SUSY corrections to the $h \rightarrow b \bar{b}$ partial decay width may nevertheless reduce the Higgs total width and hence lead to an increase of the $h \rightarrow \gamma\gamma$ branching ratio. The  presence of a light neutralino could also have a strong impact on the Higgs signal since this would increase the Higgs decay width into invisible particles \cite{Belanger:2001am,Englert:2011us,AlbornozVasquez:2011aa} and therefore decrease the signal in visible channels such as $h \rightarrow \gamma \gamma$.

In the NMSSM (a singlet extension of the MSSM \cite{Abel:1992ts}), the Higgs mass receives additional corrections from the singlet sector. As a result the fine-tuning problem is less severe and one more naturally obtains Higgs masses of about 125 GeV~\cite{Ellwanger:2009dp,Ellwanger:2011mu,Hall:2011aa}. In ~\cite{Gunion:2012zd}, it was found that the signal strength could reach at most the SM value within the framework of the  constrained NMSSM. As in the MSSM, the contribution of SUSY particles in the loop-induced $hgg$ and $h\gamma\gamma$ couplings lead to an overall suppression of the signal strength. Hence, it would be difficult to explain the central value $\sigma/\sigma_{SM}\sim 2$ reported by  both collaborations. Values of $R_{gg \gamma \gamma}$  larger than unity were nevertheless found in the NMSSM when the Higgs doublet was heavily mixed with the lightest Higgs singlet ~\cite{Ellwanger:2010nf,Ellwanger:2011aa,Kang:2012tn}. Indeed, in this case, the singlet component of the Higgs can cause a greater suppression of the partial decay width for $H_i\rightarrow b \bar{b}$ than for $H_i \rightarrow \gamma\gamma$, thus boosting the $H_i \rightarrow \gamma \gamma$ branching ratio. (Here $H_1,H_2$ are the two lightest scalar Higgses of the NMSSM.)

The results in ~\cite{Ellwanger:2010nf,Ellwanger:2011aa,Kang:2012tn} therefore raise hopes to find  Higgs candidates in the NMSSM which could fit recent LHC data. However, there is no indication of how likely these sets of parameters are with respect to the rest of the parameter space and whether they fit both the mass and signal strength when constraints on neutralino dark matter are applied. For example it may be that the LSP  in these configurations leads to a too large relic density with respect to cosmological observations, thus excluding the model. Furthermore,  direct detection limits from the XENON100 experiment~\cite{Aprile:2011hi} or indirect detection limits from gamma-rays from dwarf spheroidal galaxies~\cite{Abdo:2010ex} (which are known to constrain the properties of the neutralino LSP, especially if it is light~\cite{AlbornozVasquez:2011js}) might have an impact on Higgs properties.  

In this paper, we focus on the NMSSM and address the question of the Higgs mass and signal strength in the $\gamma\gamma$ and $WW/ZZ$ modes  when both dark matter and LHC constraints on the Higgs sector and on supersymmetric particles are simultaneously taken into account.  We will assume Friedman-Robertson-Walker (FRW) cosmology, standard thermodynamics and the freeze-out mechanism to compute the relic density.

The paper is organised as follows.
In Section II, we recall the parameter space for the NMSSM and the relevant constraints from colliders and astroparticle physics. We recall the LHC searches useful for our analysis in Section III. We determine the parameter space and give the characteristics of the Higgs which can fit the latest LHC data when neutralinos are required to be lighter than 15 GeV in Section IV and when this condition is relaxed in Section V.  We conclude in Section VI.

\section{The NMSSM and previous dark matter studies in the NMSSM}

\subsection{The NMSSM} 

The NMSSM is a simple extension of the MSSM that contains an additional gauge singlet superfield. The  VEV of this singlet induces an effective $\mu$ term which is naturally of the order of the electroweak scale, thus providing a solution to the naturalness problem~\cite{Ellwanger:2009dp}.
The model contains  three CP-even  ($H_1,H_2, H_3$) and two CP-odd ($A_1,A_2$) Higgs bosons: one more than the MSSM in both cases, due to mixing of the singlet with the two doublets.

Among the interesting features of the NMSSM, there is a coupling $\lambda S H_u H_d$ in the superpotential which leads to a positive contribution to the mass of the SM-like Higgs boson for small values of $\tan\beta$. In addition, the presence of singlet/doublet mixing  can lead to an $H_2$ Higgs with a mass within the observed range and a light eigenstate, $H_1$, compatible with all collider bounds despite the fact that $m_{H_1}$ is much smaller than $\sim$ 120 GeV. 
This happens if the singlet component of $H_1$ is large enough to reduce its couplings to the SM fields~\cite{Ellwanger:2009dp}.  When this singlet component is not large, $H_1$ becomes SM-like and $H_2$ can be much heavier than $\sim$120 GeV. In the NMSSM, it is therefore possible to either have  $H_1$ or $H_2$ in the [122-128]~GeV mass range (or even both), as favoured by the latest LHC data.

The NMSSM also contains an additional neutralino, referred to as the singlino. Due to its singlet nature the singlino can be very light, hence providing a potential light dark matter candidate.  However many of these scenarios are excluded, as shown in \cite{Vasquez:2010ru,AlbornozVasquez:2011js,Vasquez:2012px}, because they  overproduce gamma rays (or radio emission) in the galaxy or appear to be in conflict with latest direct detection experiment constraints.

\subsection{Previous scans and NMSSM parameter space}
In order to make sure that the Higgs scenarios that we consider are all relevant, we base our present analysis on \cite{Vasquez:2010ru,AlbornozVasquez:2011js,Vasquez:2012px} in which we  explored the NMSSM parameter space in light of particle physics and astroparticle physics constraints. 
In these studies, based on two different Markov Chain Monte Carlo (MCMC) analyses,  we required that the LSP relic density does not exceed the WMAP observed value (but it can be much lower than this, hence calling for another type of particles to solve the dark matter problem) and also took into account limits from B- physics, $(g-2)_\mu$, as well as LEP and Tevatron  limits on the Higgs and SUSY particles. 
These include in particular limits from the invisible width of the Z as well as from neutralino production at LEP. LHC limits on the Higgs sector computed with NMSSMTools~\cite{Ellwanger:2005dv} were also included as well as theoretical constraints on the model (Landau pole or unphysical global minimum). Additional constraints such as direct detection limits from XENON100~\cite{Aprile:2011hi}, gamma rays from dwarf spheroidal (dSph) galaxies probed by Fermi-LAT~\cite{Strigari:2006rd}  and the radio emission in the Milky Way (MW) and in galaxy clusters \cite{Boehm:2002yz,Boehm:2010kg} were superimposed on the parameter space selected by the MCMC.

In the first analyses \cite{Vasquez:2010ru,AlbornozVasquez:2011js}, we imposed the condition that the neutralino be relatively light  (between 1 and 15~GeV). This was motivated by hints of a signal in direct detection experiments~\cite{Cresst,Aalseth:2010vx,Bernabei:2010mq}.  
However since these signals have not been established yet, we relaxed this condition on the mass in the second study \cite{Vasquez:2012px}.
Even though the second study encompass all LSP masses, a separate analysis with a prior on the neutralino mass
is needed to ensure a complete coverage of the parameter space of this rather fine-tuned scenario and investigate thoroughly the special features associated with light neutralinos. For this reason we continue to keep the two analyses separate in this work.

The model that we considered in both analyses has input parameters which are defined at the weak scale. The free parameters are taken to be the gaugino masses $M_1,M_2,M_3$, the Higgs sector parameters $\mu,\tan\beta$, $\lambda,\kappa,A_\lambda,A_\kappa$, the common  soft  masses for left- and right-handed sleptons $M_{\tilde l_L}$, $M_{\tilde l,R}$ and the common soft masses for the squarks of the first and second generation ($M_{\tilde q_{1,2}}$)  and third generation ($M_{\tilde q_3}$). At last, we take one  non-zero trilinear
coupling, $A_t$. For more details, see ~\cite{Vasquez:2012px}. 

For the analysis with the $ m_{\chi} < 15$ GeV requirement, we consider only scenarios with one common soft mass for sleptons ($M_{\tilde l_L} = M_{\tilde l,R}$), one common mass for squarks ($M_{\tilde q_3} = M_{\tilde q_{1,2}}$), and the gaugino mass relation $M_2=\frac{1}{3}M_3$.  
The latter is imposed to reduce the number of free parameters knowing that the gluino does not play an important role in dark matter observables for  light neutralinos.  For the analysis {\it without} the $ m_{\chi} < 15$ GeV requirement (which we explored in the later work \cite{Vasquez:2012px}) we relaxed these three conditions, giving three further free parameters.

Details of the MCMC  (including the range of parameters for the scans and limits on observables) were given in Table I of \cite{Vasquez:2010ru} for the light neutralino case and in \cite{Vasquez:2012px} for the general case.\footnote{Since these scans were performed, a new upper limit on $B_s\rightarrow \mu^+\mu^-$ was published~\cite{Aaij:2012ac}. However we have checked {\it a posteriori} that the bulk of the points fall below this limit, thus having little impact on our analysis, see ~\cite{AlbornozVasquez:2011js} in the light neutralino case. } Physical quantities are computed by using micrOMEGAs2.4~\cite{Belanger:2010gh} that we interfaced  with NMSSMTools~\cite{Ellwanger:2005dv} for the computation of the NMSSM spectrum and particle physics constraints. 
For the case of the heavy neutralino we extended our previous analysis in order to  explore more precisely the region outlined in~\cite{Ellwanger:2005dv} where the light Higgses have a large singlet component and are heavily mixed.  For this we started some MCMC chains  in the region with
 $\lambda>0.5$ and $\tan\beta<5$.

In these scans, computing the relic density provides an important constraint on the Higgs sector. The latter was required to satisfy
$\Omega_{\rm WMAP} \, h^2 > \Omega_\chi \, h^2 > 10\% \Omega_{\rm WMAP} \, h^2$ with 
$\Omega_{\rm WMAP} \, h^2=0.1131\pm 0.0034$~\cite{Komatsu:2008hk}. 
This sets a constraint on the neutralino pair annihilation cross section in the primordial Universe. Since neutralinos lighter than 15~GeV can either be binos (as in the MSSM) or singlinos,  the easiest way to ensure significant annihilations  is through resonant exchange of a light scalar or pseudoscalar Higgs~\cite{Vasquez:2010ru}. Hence light $A_1$ or $H_1$   ($m_{H_1,A_1} < 30$ GeV) are preferred. However,  when the neutralino mass is large enough, annihilation mechanisms through Z-exchange or light sleptons can also be efficient  and a very light Higgs singlet is no longer important. 

\section{Higgs and SUSY searches at LHC used for this analysis}

For a Higgs lighter than 140~GeV as preferred by the LHC results,  the main search channel is $gg\rightarrow H\rightarrow \gamma\gamma$,
while the channel  $gg\rightarrow H\rightarrow VV$ also contributes (V denoting either a W or a Z). Therefore we compute $R_{gg\gamma\gamma}$, $R_{ggZZ}$  as well as $R_{ggWW}$, see section~\ref{sec:intro}, where the decays also include the virtual W/Z.
Note that $\sigma(gg \rightarrow H)$ is taken to be proportional to $\Gamma(H\ra gg)$ even though QCD corrections are different for the two processes. One can reasonably  assume that the effect of QCD corrections cancels out when taking the ratio of the NMSSM to the SM value. 
In what follows, we  impose up-to-date LHC constraints on the Higgs sector using {\tt HiggsBounds 3.6.1beta} \cite{Bechtle:2008jh,Bechtle:2011sb}~\footnote{Scripts for interfacing NMSSMTools with {\tt HiggsBounds} are available at \href{http://www.ippp.dur.ac.uk/~SUSY}{\bf http://www.ippp.dur.ac.uk/$\sim$SUSY}.}. In particular, the recent results in the two-photon mode are taken into account as well as the limits from the search for $H\rightarrow \tau \bar{\tau}$ which impact the heavy Higgs doublet sector of supersymmetric models~\cite{atlas:2011,cms:2011}.

To find Higgs candidates compatible with the latest LHC data, we impose (on top of all the particle physics and astroparticle constraints set in \cite{Vasquez:2010ru,Vasquez:2012px})  LHC limits on sparticles. In particular, we take into account the exclusion limit coming from the  ATLAS $1.04 \ \rm{fb}^{-1}$ search for squarks and gluinos via jets and missing $E_T$ \cite{Aad:2011ib}. For each SUSY point, signal events were generated\footnote{Note that the matrix elements used by all event generators for the hard production of two SUSY particles are accurate only to leading order in perturbative QCD. It is therefore desirable to supplement the resulting signal cross-section with an NLO K-factor for the production process, obtained in a separate calculation. In the MSSM, Prospino is commonly used; unfortunately for the NMSSM there is no automated calculation of NLO cross-sections publicly available. Exclusion calculated without the K-factor (O(1-3) in the MSSM) is therefore slightly conservative.} using  \textsf{Herwig++ 2.5.1}~\cite{Bahr:2008pv,Gieseke:2011na}. Experimental cuts of each search channel were then applied using \textsf{RIVET 1.5.2}~\cite{Buckley:2010ar}. For the ATLAS jets and missing $E_T$ searches, these are included~\cite{Grellscheid:2011ij} in the \textsf{RIVET} package.

In general these limits exclude the first and second generation squarks lighter than $0.6-1$ TeV and gluinos lighter than $\sim0.5$~TeV. However they rely on the fact that, in the constrained MSSM, there are large branching ratios of the gluinos and squarks into jets and the neutralino LSP (for the RH squarks this branching ratio is nearly 100$\%$). In the NMSSM, this is not always the case: when the LSP is purely singlino, the squarks and gluinos cannot decay to this LSP directly but must do via an intermediate particle, frequently the second-lightest neutralino. As noted in \cite{Das:2012rr} this reduces the acceptance into jets + missing $E_T$ search channels, as the extra step reduces the missing $E_T$ and may result in leptons. (SUSY searches with leptons would have in fact greater sensitivity but they do not compensate for the loss of sensitivity in the 0-lepton search \cite{Das:2012rr}). If the intermediate state decays into the LSP and a jet, there will also be greater alignment between the missing $\mathbf{p}_T$ and one of the jets -- failing the angular separation trigger  $\Delta\phi(\text{jet},\mathbf{p}_T^{\text{miss}})$.

In the analysis where we require sub-$15\,$GeV neutralinos, whenever the coloured sparticles are light enough to be within reach of the LHC they are associated with a singlino-like $\tilde\chi_1^0$. Here the jets + missing $E_T$ search is observed to be less sensitive for all of the aforementioned reasons and excludes very few of the  points\footnote{An interesting (rare) exception seen is the case of the intermediate particle decaying to the LSP and a light Higgs that decays fully invisibly -- giving large missing $E_T$.}. In the analysis with unconstrained neutralino mass, however, the singlet sector particles are generally much heavier, so that the LSP is not singlino-like. Thus the usual $\tilde{q}\rightarrow q \tilde\chi_1^0$ and $\tilde{g}\rightarrow qq \tilde\chi_1^0$ decays take place and the familiar jets + missing $E_T$ exclusion is observed: $m_{\tilde{q}}\gtrsim0.6-1$ TeV, $m_{\tilde{g}}\gtrsim0.5$ TeV.

 \begin{figure}	
\centering
\includegraphics[width=0.47\textwidth]{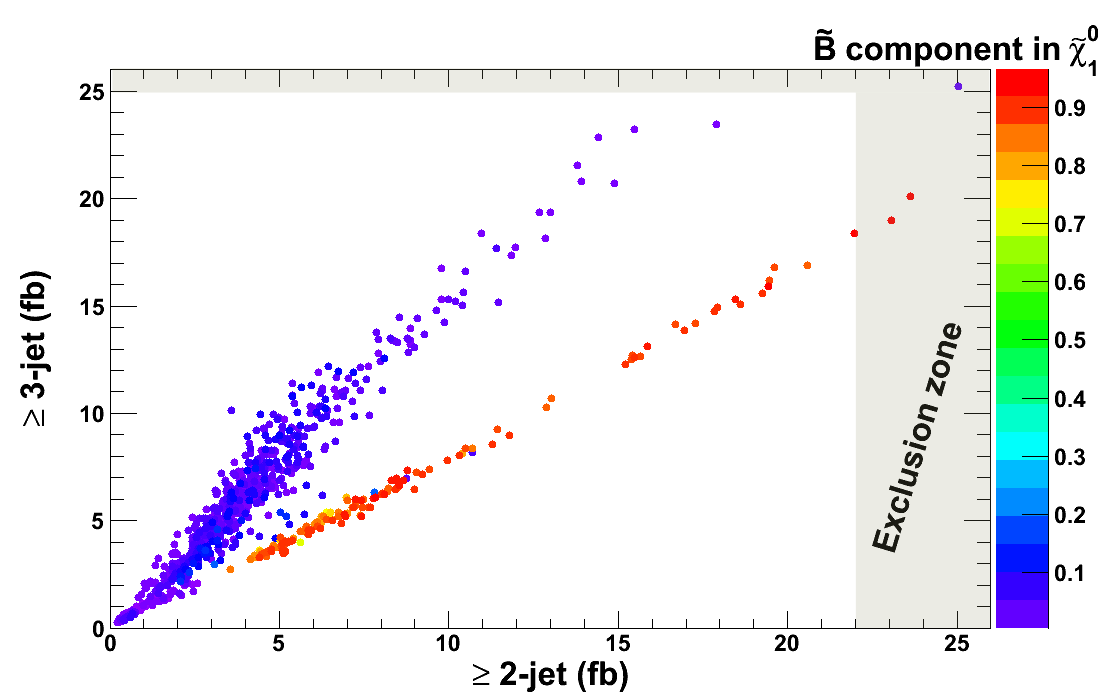}
\caption{Signal cross sections in the two- and three-jets channels of the jets + missing $E_T$ search, for points in the light neutralino MCMC scan. Red points have a bino-like LSP, blue points a singlino-like LSP. The exclusion zone ($\sigma_{2j}>22\text{fb},\:\sigma_{3j}>25\text{fb}$) is shaded.}
\label{fig:23jets}
\end{figure}

Finally, it is interesting to note that the relative acceptance into different search channels can help to distinguish a bino-like LSP from a singlino-like LSP: the latter produces a higher average number of jets in the cascade. Fig.~\ref{fig:23jets} illustrates this: we plot the signal cross sections in the two- and three-jets channels of the jets + missing $E_T$ search, showing both bino- and singlino-like LSP points from the analysis with $m_{\chi_1^0}<15$ GeV. The graph also shows the low level of exclusion -- singlino-like LSP points for the reasons discussed above, bino-like LSP points simply because the squarks are heavier and out of reach.

\section{Higgs signal strength in scenarios with neutralinos lighter than 15 GeV}

For a model to be compatible with the data, it is required that the SM-like Higgs be in the observed mass range (say  [122-128] GeV) and the signal strength be consistent with the data. As a criteria for $R_{gg\gamma\gamma}$, we choose a 2$\sigma$ error bar around the central value determined by ATLAS, thus $0.4<R_{gg\gamma\gamma}<3.6$. This is also compatible with CMS results and SM expectations (where $R_{gg\gamma\gamma}^{SM} \equiv 1$).

 \subsection{Loose constraint on the neutralino relic density}
 
\begin{figure}
	\centering
		\includegraphics[width=0.47\textwidth]{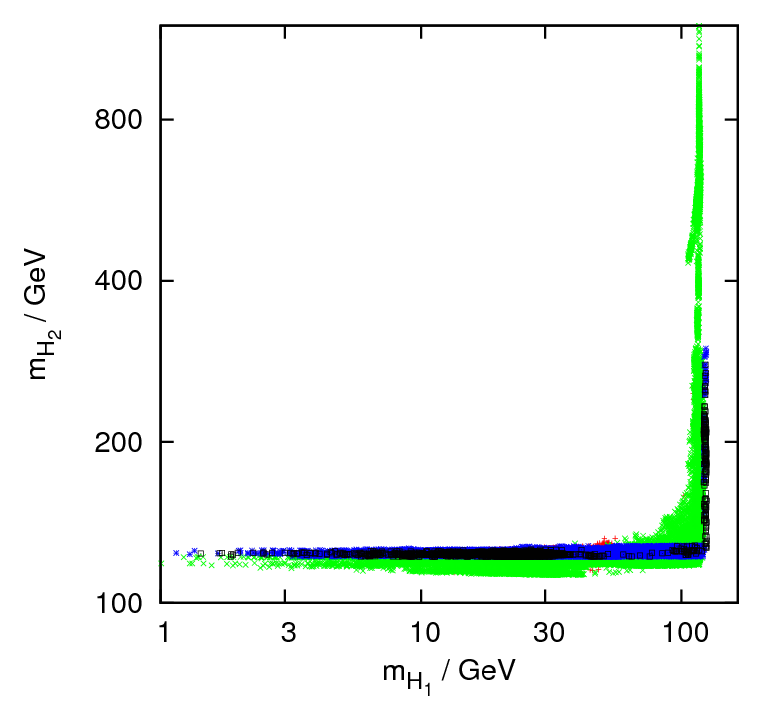}
		\includegraphics[width=0.47\textwidth]{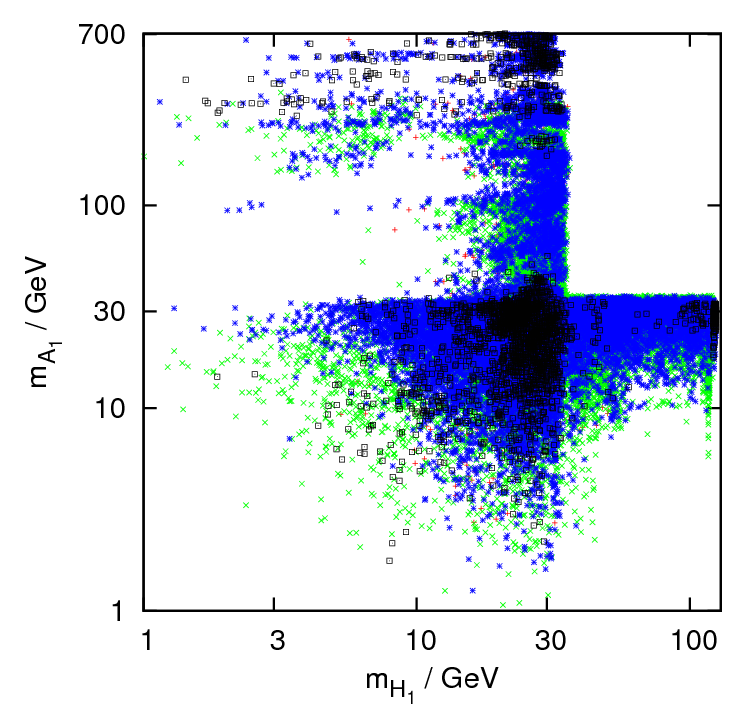}
	\caption{Masses of the Higgs scalars $H_1,H_2$ and pseudoscalar $A_1$. Red points are ruled out either by {\tt HiggsBounds} constraints or the ATLAS $1\text{fb}^{-1}$ jets and missing $E_T$ SUSY search. Green points have no Higgs with a mass in $122-128$~GeV, blue points have a Higgs ($H_1$ and/or $H_2$) within this mass range, and black points have such a Higgs with ${R_{gg\gamma\gamma}}>0.4$.}
\label{fig:mh12}
\end{figure}

In this subsection, we present the results for the Higgs sector when the neutralino mass satisfies $m_{\tilde\chi_1^0} < 15$~GeV and one applies only an upper limit on the relic density.  Imposing the condition of a light neutralino leads the MCMC to select models containing  a light Higgs ($m_{H,A} \lesssim 30$~GeV) which is mostly a singlet. 
As mentioned earlier, this could be either a scalar $H_1$ or a pseudoscalar $A_1$. As illustrated in Fig. \ref{fig:mh12},  we found that $H_1$ is typically much below the electroweak scale (thus with a large singlet component) when $H_2$ is SM-like. When $H_2$ is much heavier than 125 GeV, $H_1$ is SM-like (reaching at most up to $\sim$122~GeV) and the pseudoscalar $A_1$ is basically a light singlet, see Fig.~\ref{fig:mh12}.  It is nevertheless possible for both scalars to be heavily mixed and have a mass around 100-130~GeV (in this case $A_1$ has to be light). In Fig.~\ref{fig:mh12} blue and black points show the scenarios with at least one of the scalars within the range preferred by ATLAS and CMS. As one can see,  this is generally $H_2$ since $M_{H_1}$ barely exceeds 122~GeV.

\begin{figure}	
\centering
\includegraphics[width=0.47\textwidth]{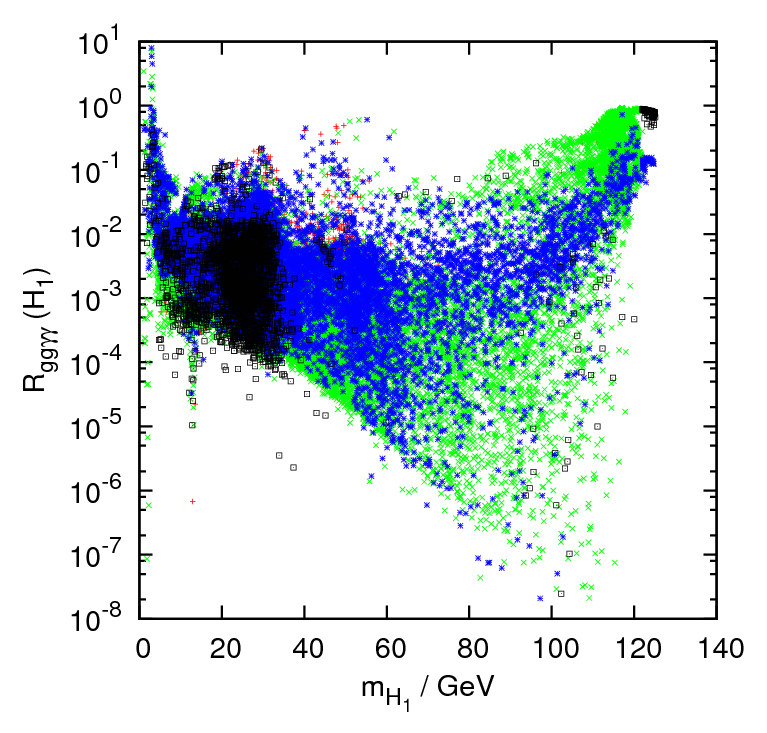}
\includegraphics[width=0.47\textwidth]{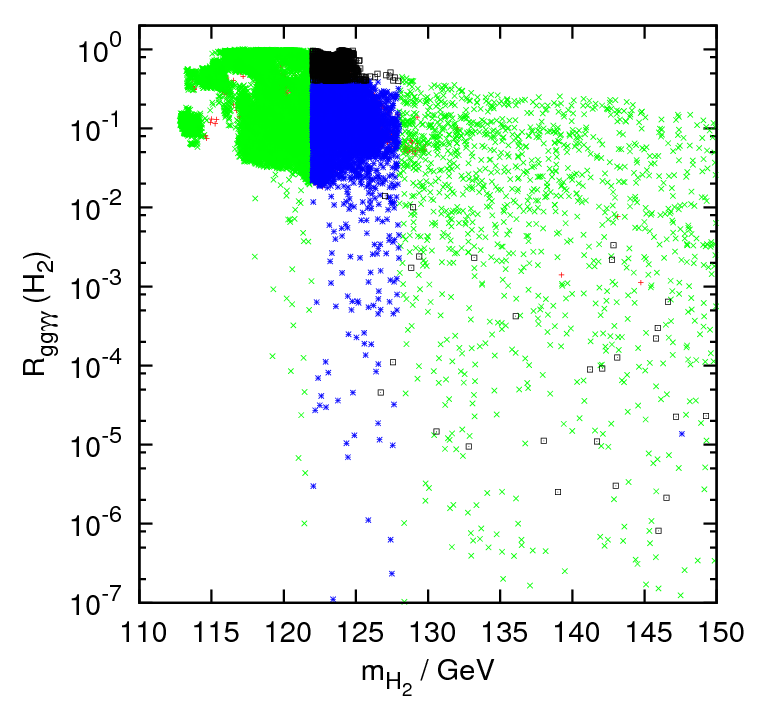}
\caption{$R_{gg\gamma\gamma}$ as a function of the mass of $H_1$ (top panel) and of $H_2$ (the more usual candidate; bottom panel) in the light neutralino LSP model. Same colour code as Fig.~\ref{fig:mh12}.
 }
\label{fig:gggg_nmssm}
\end{figure}

The predictions for $R_{gg\gamma\gamma}$  as a function of the $H_2$ mass are displayed in Fig.~\ref{fig:gggg_nmssm} (bottom panel). We only display the region where this channel is relevant, that is when $m_{H_2} < 150$ GeV. 
As one can see, all the configurations that were selected by the MCMC (and which were compatible with the constraints that we imposed for the scans) have $R_{gg\gamma\gamma} < 1 $.  An explanation is that, although $H_2$ couplings are usually SM-like,  large suppressions in $R_{gg\gamma\gamma}$ are possible because the width of the Higgs is enhanced by many new non-standard decay channels~\cite{Cao:2011re}. In particular, $H_2$ can decay into two neutralinos, two light scalar Higgses ($H_1$) or two light pseudoscalar Higgses ($A_1$) which reduces significantly the branching ratio into two-photons.

In Fig.~\ref{fig:gggg_nmssm}, all the points which do not satisfy either the newest {\tt HiggsBounds} limits or the SUSY searches in jets plus missing $E_T$ are colored in red. As mentioned above very few of these points are excluded by SUSY searches. The points which fall within the Higgs observed mass range are highlighted in blue. Scenarios where the strength of the signal in $\gamma\gamma$ is also compatible with the 2$\sigma$ range reported by ATLAS ($R_{gg\gamma\gamma}>0.4$) 
are represented by black squares.

We have also computed $R_{gg\gamma\gamma}$ for $H_1$ and found that it is usually much below unity because $H_1$ has a large singlet component. Only a few points have  $R_{gg\gamma\gamma}\approx 1$ and they correspond to either a SM-like $H_1$ with a mass near 122~GeV or to a very light singlet.  
 
Let us now examine the predictions for $R_{ggVV}$ (with $V=W$ or $V=Z$) as a function of the $H_2$ mass. 
Again the maximum value is unity. $R_{ggVV}$ can be much suppressed especially because of a large branching ratio into invisibles,  hence a correlation between the large suppression in the $VV$ channel and in the $\gamma\gamma$ channel.  
The results for $R_{ggVV}$ are very similar to those for $R_{gg\gamma\gamma}$ shown in Fig.~\ref{fig:gggg_nmssm}.
It is interesting to note that for all  the scenarios where the $H_2$ mass is above 130~GeV (i.e. where the WW/ZZ modes are the dominant decay channels), the signal strength $R_{ggVV}$ is suppressed by  at least  a factor 3 and even often suppressed by one order of magnitude. Hence, a non SM-like $H_2$ Higgs with a mass of about 140 GeV (corresponding to a scenario similar to the MSSM) is still allowed.  

\begin{figure}
	\centering
		\includegraphics[width=0.47\textwidth]{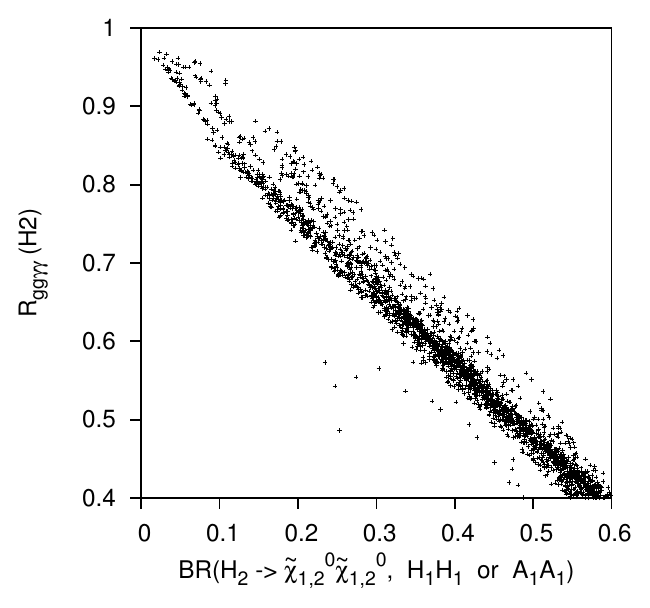}
	\caption{Showing the reduction in diphoton signal strength from $H\rightarrow\gamma\gamma$ competing with new BSM decays. Here only points with a Higgs mass between 122-128~GeV and with $R_{gg\gamma\gamma}>0.4$ are included.}
\label{fig:SignalStrength_V_BSMdecays}
\end{figure}

\begin{figure}
	\centering
		\includegraphics[width=0.25\textwidth]{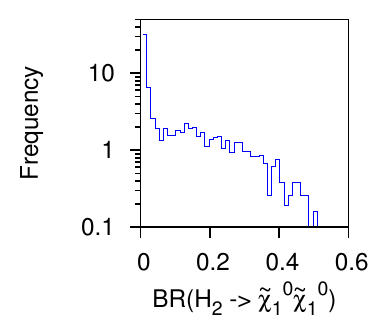} \hspace*{-0.03\textwidth}
		\includegraphics[width=0.25\textwidth]{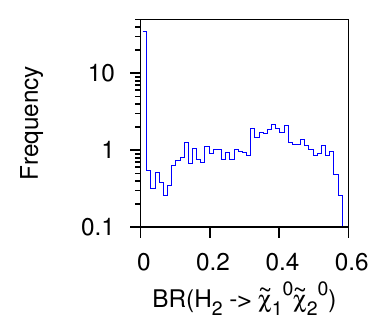}
                \includegraphics[width=0.25\textwidth]{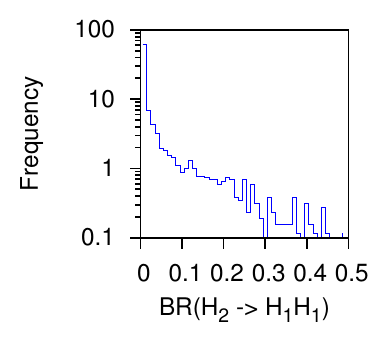} \hspace*{-0.03\textwidth}
		\includegraphics[width=0.25\textwidth]{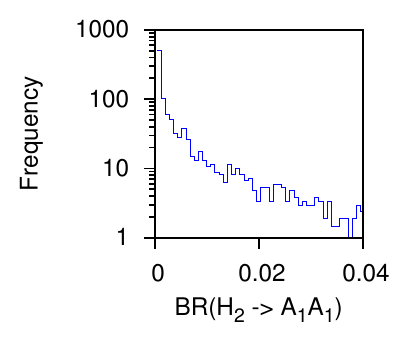}
	\caption{The distribution of the different BSM decay channels for the same points as in Fig. \ref{fig:SignalStrength_V_BSMdecays} (where $H_2\rightarrow\tilde\chi_2^0\tilde\chi_2^0$ is negligible). Histograms have unit area.}
\label{fig:distrib}
\end{figure}

Interestingly enough, these NMSSM scenarios could not be distinguished from the MSSM based on the two-photon channel (and/or the WW channel) since both have a maximum signal strength of about unity. However a search for the singlet Higgs would give a distinctive NMSSM signature. As we have mentioned, these light neutralino models must also have a light singlet Higgs and thus non-standard decays occur. The effect of these decays on the signal strength is shown in Fig.~\ref{fig:SignalStrength_V_BSMdecays} and the distribution of the decays in Fig.~\ref{fig:distrib} for the points compatible with the observed excess. 
Clearly,  too large branching ratios into non standard modes (such as $\tilde\chi_1^0 \tilde\chi_i^0, H_1 H_1, A_1 A_1$) would render the two-photon mode invisible or suppressed with respect to the SM prediction~\cite{Cao:2011re}. In fact, in order for the signal strength  $R_{gg\gamma\gamma} $ to be compatible with $R_{gg\gamma\gamma} > 0.4$,  the branching ratio  $BR_{H_2 \rightarrow \rm{invisible}}$ must be lower than $\sim60\%$.

The existence of decay modes such as $H_2\rightarrow H_1 H_1$ or $ A_1 A_1$, with the singlet Higgs further decaying into Standard Model particles remains nevertheless interesting because such modes give a distinctive signature which could be searched for at LHC and would constitute evidence for new physics if they are found. Extraction of this signal from background via jet substructure techniques has been studied for $H \rightarrow 2 A \rightarrow 4\,  \tau$ in \cite{Englert:2011iz}, $H\rightarrow \, 2 \, A \, \rightarrow \tau \, \bar{\tau} \, \mu \, \bar{\mu}$ in \cite{Lisanti:2009uy}, and $H\rightarrow 2 A \rightarrow 4 \ g$ (less relevant for SUSY due to $\tan\beta$ suppression) in \cite{Falkowski:2010hi,Chen:2010wk}. These decays with an intermediate scalar ($H_1$) instead of an intermediate pseudoscalar ($A_1$) give the same signal.

A reduced two-photon signal could also be due to a sizable branching ratio of the Higgs into $\tilde\chi_1^0\tilde\chi_1^0$ and $\tilde\chi_1^0\tilde\chi_2^0$. Indeed this generally dominates over the decay to lighter Higgses in this sample. $H\ra\tilde\chi_1^0\tilde\chi_2^0$ occurs mainly when the LSP is singlino: the bino/higgsino/singlino NLSP can be significantly lighter than 100~GeV thus allowing sufficient phase space for the decay. This mode is also possible  for a bino LSP and a higgsino NLSP but the phase space is quite limited. Note that the NLSP $\tilde\chi_2^0$ can also have decay modes into light Higgses which further decay in fermion pairs. 

Another distinctive feature of the NMSSM Higgs sector would be the direct search for $H_1$. We have seen that the $gg\rightarrow H_1\rightarrow \gamma\gamma$ channel is suppressed. This is also true for other channels; indeed the couplings of $H_1$ to SM particles  is suppressed by the singlet component. For the $b\bar{b}$ production mode, the suppression can be in part compensated by a $\tan\beta$ enhancement. However in our scans we found that the $H_1bb$ coupling could reach at most its SM value. Thus the production of  $H_1$ in association with b-quarks followed by the decay of the Higgs into tau pairs does not benefit from an enhancement over the SM expectations. 

\subsection{Upper and lower limit on the neutralino relic density}
\label{sec:wmap}

\begin{figure}	
\centering
\includegraphics[width=0.47\textwidth]{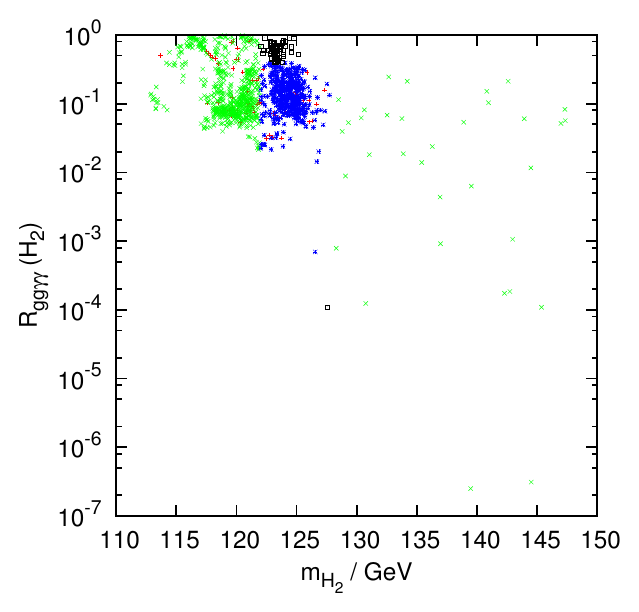}
\caption{$R_{gg\gamma\gamma}$ as a function of $M_{H_2}$ in the light neutralino LSP model  when the relic density is compatible with WMAP. Same colour code as Fig.~\ref{fig:mh12}. }
\label{fig:gggg_wmap}
\end{figure}

If we now select only the points which predict a relic density at the WMAP observed value (more precisely we impose $ \Omega_{WMAP}^{max} h^2> \Omega h^2>0.999 \ \Omega^{min}_{WMAP} h^2$ where $\Omega^{max,min}_{WMAP}$ is $\pm 1 \sigma$ from the central measured value of WMAP), we find strong constraints on the parameter space as light Higgses in the s-channel or in the final states may reduce the relic density too much. However,  the results for the Higgs sector which are compatible with the latest LHC data are similar to the ones that were discussed in the previous subsection. 
For completeness we show the expectations for $R_{gg\gamma\gamma}$ for these points in Fig.~\ref{fig:gggg_wmap}.

\section{Signal strength in scenarios with heavy neutralinos}

We repeated the analysis for the MCMC scan in which there is no $\mlsp < 15$ GeV requirement. We first consider the impact of Higgs searches at LHC. The effect of also imposing the relic density condition and SUSY searches will be discussed in section~\ref{sec:heavy_susy}.

\subsection{Loose constraint on the relic density}

\begin{figure}	
\centering
\includegraphics[width=0.47\textwidth]{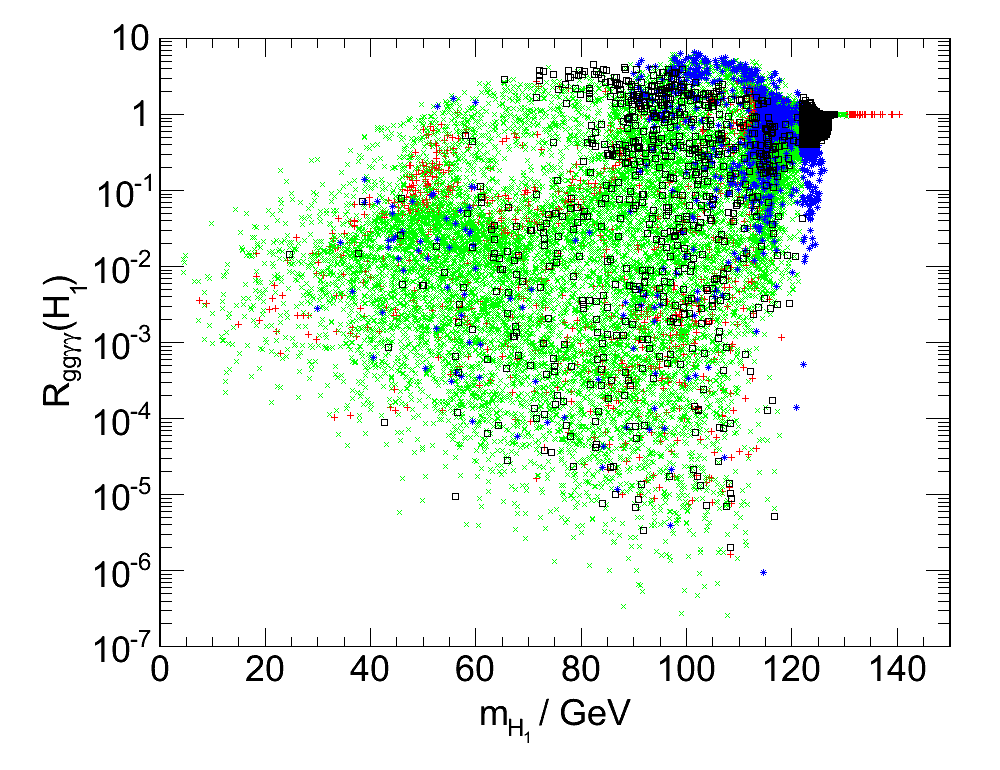}
\includegraphics[width=0.47\textwidth]{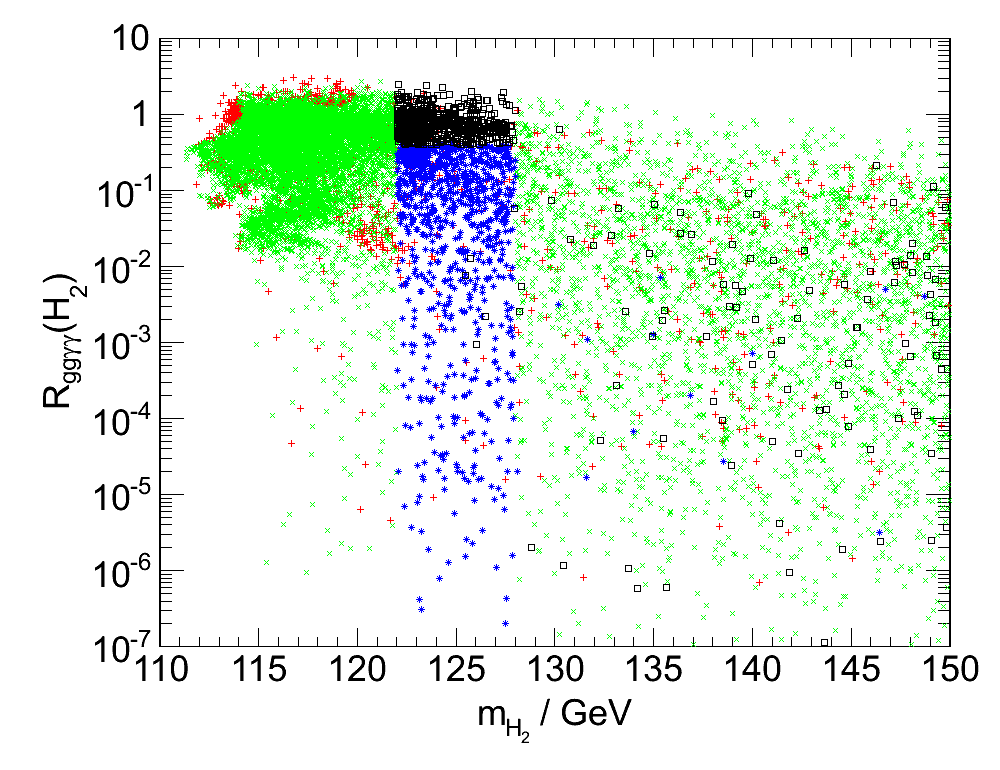}
\caption{$R_{gg\gamma\gamma}$ as a function of the mass of $H_1$ (top panel) and of $H_2$ (bottom panel) in the arbitrary neutralino LSP model. In the top panel most points are on top of one another at high mass and high $R_{gg\gamma\gamma}$.
Same colour code as Fig.~\ref{fig:mh12} except red points are denoting only exclusion by {\tt HiggsBounds}.}
\label{fig:gggg_all}
\end{figure}

As opposed to the previous case with a light neutralino we find that in general the Higgs in the mass region preferred by the LHC is a SM-like $H_1$. Indeed, without very light neutralinos, a very light singlet sector is not needed for resonant annihilations. Thus the associated values for $R_{gg\gamma\gamma}$ are naturally of order unity (see the black squares in Fig.~\ref{fig:gggg_all}, top panel).  Nevertheless we found cases where $R_{gg\gamma\gamma}<0.4$, when invisible decay modes (such as $H_1 \rightarrow \tilde\chi_1^0\tilde\chi_1^0$ or $H_1 \rightarrow A_1 A_1$)  are kinematically accessible.  

We also found points with $R_{gg\gamma\gamma}>1$; these were  mostly associated with  a Higgs mass below the LHC preferred range.
Indeed, the light Higgs has a  large singlet component, which leads to suppressed couplings to SM particles, in particular to $b\bar{b}$.
The suppression of  the partial decay width for $H_1\rightarrow b\bar{b}$, and thus of the total width, leads to  a larger branching ratio in the $H_1\rightarrow \gamma\gamma$~\cite{Ellwanger:2010nf,Ellwanger:2011aa,Cao:2012fz}. For $m_{H_1}>122$~GeV,  the increase in $R_{gg\gamma\gamma}$ is generally modest (below 20\%) 
except for a few points where both Higgses are heavily mixed and have a large singlet component. 
An enhancement with respect  to the SM expectations is found  for large values of $\lambda$ and small values of  $\mu$ (below $\approx 200$~GeV). Under these conditions the mixing between the singlet and doublet Higgs is large and the singlet Higgs is light. Recall that the mass of the singlet Higgs depends on $\mu$ with $m_S^2=\kappa \mu/\lambda (A_\kappa +4 \kappa\mu/\lambda)$ ~\cite{Belanger:2008nt,Ellwanger:2009dp}.

Note that in Fig.~\ref{fig:gggg_all} top panel, we found scenarios where $H_2$ is in the observed range while $H_1$ is lighter and has $R_{gg\gamma\gamma}\gg 1$. This  means that even though $H_1$ has evaded present constraints from LHC, these points offer good prospects for discovery of a second Higgs scalar in the next run of the LHC.  Such a signal would allow to distinguish the NMSSM Higgs sector from the MSSM one.

$H_2$ masses extend over a wide range (all the way to several TeV's) and include some points in the mass region preferred by the LHC. The values of $R_{gg\gamma\gamma}$ for $H_2$ are displayed in Fig.~\ref{fig:gggg_all} (bottom panel)  for the  range of  masses where  the two-photon search mode is relevant. We found that the signal strength reaches values as high as  $R_{gg\gamma\gamma}=2$.
This enhancement with respect to the SM expectations is found when $H_2$ has some singlet component and a suppressed partial width to $b\bar{b}$. ($H_1$, conversely, has an enhanced $b\bar{b}$ partial width and a reduced signal strength $R_{gg\gamma\gamma}$.  ) 

As mentioned above, when  $0.4<R_{gg\gamma\gamma}<1$ for $H_2$, the signal strength for the lighter Higgs can be enhanced.   
We have also found points where $R_{gg\gamma\gamma}<1$ for both $H_1$ and $H_2$. This can be due to the presence of invisible modes or to the presence of a singlet component for which the suppression in the Higgs coupling to $gg$ compensates the increase of the $\gamma\gamma$ branching ratio.

\subsection{With a lower limit on the relic density}
\label{sec:heavy_susy}

\begin{figure}	
\centering
\includegraphics[width=0.47\textwidth]{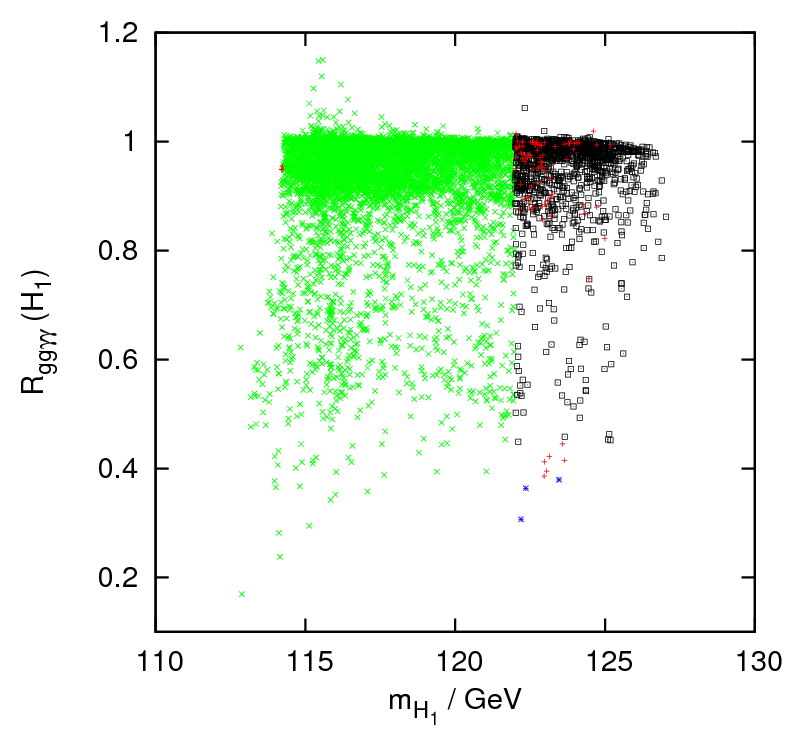}
\caption{$R_{gg\gamma\gamma}$ as a function of the mass of $H_1$ (the more usual candidate)  for points with a relic density within the WMAP range. A handful of points are also seen at masses lower than those shown, with reduced signal strength.
Same colour code as Fig.~\ref{fig:mh12}.}
\label{fig:gggg_heavy}
\end{figure}

Finally we analyse the impact of requiring the relic density condition. Imposing a strict lower limit on the relic density strongly constrains the scenarios where the singlet component of the light Higgs  leads to $R_{gg\gamma\gamma}>1$. As mentioned above, these points were found for $\mu<200$~GeV. Hence the LSP has a non-negligible higgsino component, usually leading to a relic density below the preferred WMAP range. 
After imposing the strict lower bound on the relic density, we found that $H_1$ was in general within the LHC preferred range, while $H_2$ was heavy, its mass extending above the TeV scale. $R_{gg\gamma\gamma}$ for $H_1$ is displayed in Fig.~\ref{fig:gggg_heavy}: the relic density constraint has removed most of the points where the signal strength was enhanced (the maximum value is now 1.06). 

We also considered the constraint from the jets and missing $E_T$ SUSY search for those points with a Higgs of mass $122-128~{\rm GeV}$. Exclusion is observed in this mass range for any signal strength, showing as expected that there is no direct correlation between the first and second generation squarks and the Higgs sector. 
Finally we have checked that these points were compatible with the latest upper limit on $Br(B_s\rightarrow \mu^+\mu^-)<4.5\times 10^{-9}$ from LHCb~\cite{LHCb}.

\section{Prospects and conclusion}

In this paper, we have investigated the Higgs signal strength expected in scans of the NMSSM scenarios in which we have applied particle physics as well as astroparticle physics constraints. In particular we have imposed that the neutralino relic density does not exceed the WMAP observed value nor the limit imposed by direct detection experiments  and does not overproduce gamma rays and radio waves in the galaxy. We also took into account limits from B- physics, $(g-2)_{\mu}$, as well as LEP, Tevatron and LHC limits on the Higgs and SUSY particles.

We found many scenarios where either $H_1$ or $H_2$ had a mass in the range [122-128] GeV and 
a signal strength compatible with the Standard Model. We also found scenarios where
the signal strength in the two-photon mode was as large as the excess reported by ATLAS and CMS. However for most of these points the neutralino would form only a fraction of the observed dark matter. 
If a Higgs is confirmed at the LHC with a signal strength  in excess of the SM expectations, it would be a clear indication
of BSM physics, while if the signal strength is compatible  with  SM expectations, it might be necessary to search for light pseudoscalar or scalar Higgs  in order to determine whether one has discovered the SM Higgs or BSM physics.

When insisting on a light neutralino, we found that  the most promising configurations favour a SM-like $H_2$ rather than a $H_1$ SM-like Higgs and therefore predict the existence of a light Higgs dominantly singlet. The possibility of observing a second light Higgs provides a distinct signature of the NMSSM Higgs sector. This could be done directly, e.g. via the diphoton search for an SM-like Higgs with higher luminosities; or indirectly via the decay of the heavier Higgs, e.g. $H_2\rightarrow (2 H_1\; \rm{or} \; 2A_1) \rightarrow\tau \bar{\tau}$.

\begin{figure}	
\vspace*{6mm}
\centering
\includegraphics[width=0.47\textwidth]{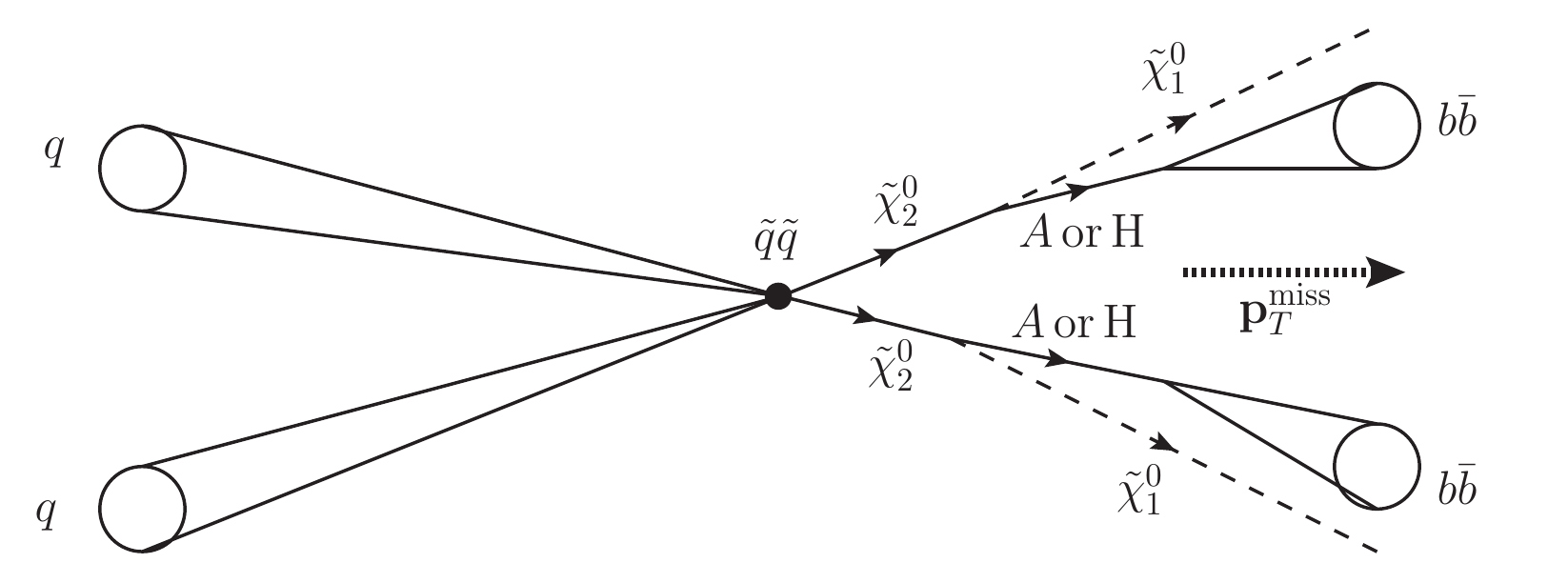}
\caption{The modified jets + missing $E_T$ signal in the NMSSM with a singlino-like LSP, when a decay $\tilde\chi^0_2 \rightarrow \tilde\chi^0_1 + {\rm Higgs}$ is kinematically accessible. Because the $\tilde\chi^0_2$ are boosted, the missing $\mathbf{p}_T$ vector will point in between the two $b$-jets.}
\label{fig:fatjets}
\end{figure}

Furthermore we note that the traditional jets + missing $E_T$ signature of squarks and gluinos in the MSSM can be modified in a very interesting way in the NMSSM, when the LSP is singlino-like and the second-lightest neutralino $\tilde\chi_2^0$ can decay to a Higgs. The $\tilde\chi_2^0$ is expected to be boosted (generically being much lighter than the squarks and gluinos), and the lighter Higgs which it decays to ($H_1$ or $A_1$) is expected to further decay into $b\bar{b}$ \footnote{Higgs $\rightarrow \tau \bar{\tau}$ would be dominant for $m_{H,A}<2m_b$}. This extra step in this cascade compared to the MSSM $2\tilde{q}\rightarrow 2q+2\tilde\chi^0_2$ adds jets and halves the missing $E_T$, dimming discovery prospects. However a distinctive topology results: the boosted nature of the intermediate $\tilde\chi_2^0$ means the missing $\mathbf{p}_T$ vector will point in between the two doubly-tagged $b$-jets. We illustrate this scenario in Fig. \ref{fig:fatjets} for disquark production. Gluinos tend to decay to a neutralino plus two jets rather than one, as is well known; otherwise the signal remains the same as for squarks. Calculation of the Standard Model background and estimation of the systematic error in measuring such a signal are clearly vital, but beyond the scope of this work.

The NMSSM Higgs sector can however be very similar to the MSSM, or even the SM, with only a light SM-like scalar and much heavier scalars and pseudoscalars. In this case searches for the superpartners, and in particular peculiar signatures associated with the singlino LSP (e.g. bino $\rightarrow$ singlino + soft $l\bar{l}$ ~\cite{Kraml:2008zr}), offer the only possibility to identify the NMSSM.

\section{Acknowledgments}
GB and CW thank the LPSC for its hospitality. The authors would like to thank
Alexander Pukhov, Ulrich Ellwanger, Michael Spannowsky and David Grellscheid
for useful discussions; also Tim Stefaniak and Oscar St\aa{}l for communications concerning  {\tt HiggsBounds}. This work was partly supported by the STFC and the CNRS-PICS grant ``Propagation of low energy positrons'' held by CB.
JD acknowledges the CMIRA 2011 EXPLO'RA DOC program of the French region Rh\^one-Alpes.


\newpage

\begin{thebibliography}{64}
\expandafter\ifx\csname natexlab\endcsname\relax\def\natexlab#1{#1}\fi
\expandafter\ifx\csname bibnamefont\endcsname\relax
  \def\bibnamefont#1{#1}\fi
\expandafter\ifx\csname bibfnamefont\endcsname\relax
  \def\bibfnamefont#1{#1}\fi
\expandafter\ifx\csname citenamefont\endcsname\relax
  \def\citenamefont#1{#1}\fi
\expandafter\ifx\csname url\endcsname\relax
  \def\url#1{\texttt{#1}}\fi
\expandafter\ifx\csname urlprefix\endcsname\relax\def\urlprefix{URL }\fi
\providecommand{\bibinfo}[2]{#2}
\providecommand{\eprint}[2][]{\url{#2}}

\bibitem[{\citenamefont{Aad et~al.}(2012{\natexlab{a}})}]{ATLAS:2012ae}
\bibinfo{author}{\bibfnamefont{G.}~\bibnamefont{Aad}} \bibnamefont{et~al.}
  (\bibinfo{collaboration}{ATLAS Collaboration}), \bibinfo{journal}{Phys.Lett.}
  \textbf{\bibinfo{volume}{B710}}, \bibinfo{pages}{49}
  (\bibinfo{year}{2012}{\natexlab{a}}), \eprint{1202.1408}.

\bibitem[{\citenamefont{Chatrchyan
  et~al.}(2012{\natexlab{a}})}]{Chatrchyan:2012tx}
\bibinfo{author}{\bibfnamefont{S.}~\bibnamefont{Chatrchyan}}
  \bibnamefont{et~al.} (\bibinfo{collaboration}{CMS Collaboration})
  (\bibinfo{year}{2012}{\natexlab{a}}), \eprint{1202.1488}.

\bibitem[{\citenamefont{Aad et~al.}(2012{\natexlab{b}})}]{ATLAS:2012ad}
\bibinfo{author}{\bibfnamefont{G.}~\bibnamefont{Aad}} \bibnamefont{et~al.}
  (\bibinfo{collaboration}{ATLAS Collaboration}),
  \bibinfo{journal}{Phys.Rev.Lett.} \textbf{\bibinfo{volume}{108}},
  \bibinfo{pages}{111803} (\bibinfo{year}{2012}{\natexlab{b}}),
  \eprint{1202.1414}.

\bibitem[{\citenamefont{Chatrchyan
  et~al.}(2012{\natexlab{b}})}]{Chatrchyan:2012tw}
\bibinfo{author}{\bibfnamefont{S.}~\bibnamefont{Chatrchyan}}
  \bibnamefont{et~al.} (\bibinfo{collaboration}{CMS Collaboration})
  (\bibinfo{year}{2012}{\natexlab{b}}), \eprint{1202.1487}.

\bibitem[{\citenamefont{Baer et~al.}(2011)\citenamefont{Baer, Barger, and
  Mustafayev}}]{Baer:2011ab}
\bibinfo{author}{\bibfnamefont{H.}~\bibnamefont{Baer}},
  \bibinfo{author}{\bibfnamefont{V.}~\bibnamefont{Barger}}, \bibnamefont{and}
  \bibinfo{author}{\bibfnamefont{A.}~\bibnamefont{Mustafayev}}
  (\bibinfo{year}{2011}), \eprint{1112.3017}.

\bibitem[{\citenamefont{Heinemeyer et~al.}(2011)\citenamefont{Heinemeyer, Stal,
  and Weiglein}}]{Heinemeyer:2011aa}
\bibinfo{author}{\bibfnamefont{S.}~\bibnamefont{Heinemeyer}},
  \bibinfo{author}{\bibfnamefont{O.}~\bibnamefont{Stal}}, \bibnamefont{and}
  \bibinfo{author}{\bibfnamefont{G.}~\bibnamefont{Weiglein}}
  (\bibinfo{year}{2011}), \eprint{1112.3026}.

\bibitem[{\citenamefont{Carena et~al.}(2011)\citenamefont{Carena, Gori, Shah,
  and Wagner}}]{Carena:2011aa}
\bibinfo{author}{\bibfnamefont{M.}~\bibnamefont{Carena}},
  \bibinfo{author}{\bibfnamefont{S.}~\bibnamefont{Gori}},
  \bibinfo{author}{\bibfnamefont{N.~R.} \bibnamefont{Shah}}, \bibnamefont{and}
  \bibinfo{author}{\bibfnamefont{C.~E.} \bibnamefont{Wagner}}
  (\bibinfo{year}{2011}), \eprint{1112.3336}.

\bibitem[{\citenamefont{Arbey et~al.}(2011{\natexlab{a}})\citenamefont{Arbey,
  Battaglia, Djouadi, Mahmoudi, and Quevillon}}]{Arbey:2011ab}
\bibinfo{author}{\bibfnamefont{A.}~\bibnamefont{Arbey}},
  \bibinfo{author}{\bibfnamefont{M.}~\bibnamefont{Battaglia}},
  \bibinfo{author}{\bibfnamefont{A.}~\bibnamefont{Djouadi}},
  \bibinfo{author}{\bibfnamefont{F.}~\bibnamefont{Mahmoudi}}, \bibnamefont{and}
  \bibinfo{author}{\bibfnamefont{J.}~\bibnamefont{Quevillon}}
  (\bibinfo{year}{2011}{\natexlab{a}}), \eprint{1112.3028}.

\bibitem[{\citenamefont{Olive}(2012)}]{Olive:2012it}
\bibinfo{author}{\bibfnamefont{K.~A.} \bibnamefont{Olive}}
  (\bibinfo{year}{2012}), \eprint{1202.2324}.

\bibitem[{\citenamefont{Espinosa et~al.}(2012)\citenamefont{Espinosa, Grojean,
  Muhlleitner, and Trott}}]{Espinosa:2012ir}
\bibinfo{author}{\bibfnamefont{J.}~\bibnamefont{Espinosa}},
  \bibinfo{author}{\bibfnamefont{C.}~\bibnamefont{Grojean}},
  \bibinfo{author}{\bibfnamefont{M.}~\bibnamefont{Muhlleitner}},
  \bibnamefont{and} \bibinfo{author}{\bibfnamefont{M.}~\bibnamefont{Trott}}
  (\bibinfo{year}{2012}), \eprint{1202.3697}.

\bibitem[{\citenamefont{Carmi et~al.}(2012)\citenamefont{Carmi, Falkowski,
  Kuflik, and Volansky}}]{Carmi:2012yp}
\bibinfo{author}{\bibfnamefont{D.}~\bibnamefont{Carmi}},
  \bibinfo{author}{\bibfnamefont{A.}~\bibnamefont{Falkowski}},
  \bibinfo{author}{\bibfnamefont{E.}~\bibnamefont{Kuflik}}, \bibnamefont{and}
  \bibinfo{author}{\bibfnamefont{T.}~\bibnamefont{Volansky}}
  (\bibinfo{year}{2012}), \eprint{1202.3144}.

\bibitem[{\citenamefont{Azatov et~al.}(2012)\citenamefont{Azatov, Contino, and
  Galloway}}]{Azatov:2012bz}
\bibinfo{author}{\bibfnamefont{A.}~\bibnamefont{Azatov}},
  \bibinfo{author}{\bibfnamefont{R.}~\bibnamefont{Contino}}, \bibnamefont{and}
  \bibinfo{author}{\bibfnamefont{J.}~\bibnamefont{Galloway}}
  (\bibinfo{year}{2012}), \eprint{1202.3415}.

\bibitem[{\citenamefont{Baer et~al.}(2012)\citenamefont{Baer, Barger, and
  Mustafayev}}]{Baer:2012uy}
\bibinfo{author}{\bibfnamefont{H.}~\bibnamefont{Baer}},
  \bibinfo{author}{\bibfnamefont{V.}~\bibnamefont{Barger}}, \bibnamefont{and}
  \bibinfo{author}{\bibfnamefont{A.}~\bibnamefont{Mustafayev}}
  (\bibinfo{year}{2012}), \eprint{1202.4038}.

\bibitem[{\citenamefont{King et~al.}(2012)\citenamefont{King, Muhlleitner, and
  Nevzorov}}]{King:2012is}
\bibinfo{author}{\bibfnamefont{S.}~\bibnamefont{King}},
  \bibinfo{author}{\bibfnamefont{M.}~\bibnamefont{Muhlleitner}},
  \bibnamefont{and} \bibinfo{author}{\bibfnamefont{R.}~\bibnamefont{Nevzorov}}
  (\bibinfo{year}{2012}), \eprint{1201.2671}.

\bibitem[{\citenamefont{Gunion et~al.}(2012)\citenamefont{Gunion, Jiang, and
  Kraml}}]{Gunion:2012zd}
\bibinfo{author}{\bibfnamefont{J.~F.} \bibnamefont{Gunion}},
  \bibinfo{author}{\bibfnamefont{Y.}~\bibnamefont{Jiang}}, \bibnamefont{and}
  \bibinfo{author}{\bibfnamefont{S.}~\bibnamefont{Kraml}}
  (\bibinfo{year}{2012}), \eprint{1201.0982}.

\bibitem[{\citenamefont{Brooijmans et~al.}(2012)\citenamefont{Brooijmans,
  Gripaios, Moortgat, Santiago, Skands et~al.}}]{Brooijmans:2012yi}
\bibinfo{author}{\bibfnamefont{G.}~\bibnamefont{Brooijmans}},
  \bibinfo{author}{\bibfnamefont{B.}~\bibnamefont{Gripaios}},
  \bibinfo{author}{\bibfnamefont{F.}~\bibnamefont{Moortgat}},
  \bibinfo{author}{\bibfnamefont{J.}~\bibnamefont{Santiago}},
  \bibinfo{author}{\bibfnamefont{P.}~\bibnamefont{Skands}},
  \bibnamefont{et~al.} (\bibinfo{year}{2012}), \eprint{1203.1488}.

\bibitem[{\citenamefont{Cao et~al.}(2011{\natexlab{a}})\citenamefont{Cao, Heng,
  Li, and Yang}}]{Cao:2011sn}
\bibinfo{author}{\bibfnamefont{J.}~\bibnamefont{Cao}},
  \bibinfo{author}{\bibfnamefont{Z.}~\bibnamefont{Heng}},
  \bibinfo{author}{\bibfnamefont{D.}~\bibnamefont{Li}}, \bibnamefont{and}
  \bibinfo{author}{\bibfnamefont{J.~M.} \bibnamefont{Yang}}
  (\bibinfo{year}{2011}{\natexlab{a}}), \eprint{1112.4391}.

\bibitem[{\citenamefont{Christensen et~al.}(2012)\citenamefont{Christensen,
  Han, and Su}}]{Christensen:2012ei}
\bibinfo{author}{\bibfnamefont{N.~D.} \bibnamefont{Christensen}},
  \bibinfo{author}{\bibfnamefont{T.}~\bibnamefont{Han}}, \bibnamefont{and}
  \bibinfo{author}{\bibfnamefont{S.}~\bibnamefont{Su}} (\bibinfo{year}{2012}),
  \bibinfo{note}{34 pages, 41 figures}, \eprint{1203.3207}.

\bibitem[{\citenamefont{Desai et~al.}(2012)\citenamefont{Desai, Mukhopadhyaya,
  and Niyogi}}]{Desai:2012qy}
\bibinfo{author}{\bibfnamefont{N.}~\bibnamefont{Desai}},
  \bibinfo{author}{\bibfnamefont{B.}~\bibnamefont{Mukhopadhyaya}},
  \bibnamefont{and} \bibinfo{author}{\bibfnamefont{S.}~\bibnamefont{Niyogi}}
  (\bibinfo{year}{2012}), \eprint{1202.5190}.

\bibitem[{\citenamefont{Djouadi}(2008)}]{Djouadi:2005gj}
\bibinfo{author}{\bibfnamefont{A.}~\bibnamefont{Djouadi}},
  \bibinfo{journal}{Phys.Rept.} \textbf{\bibinfo{volume}{459}},
  \bibinfo{pages}{1} (\bibinfo{year}{2008}), \eprint{hep-ph/0503173}.

\bibitem[{\citenamefont{Hall et~al.}(2011)\citenamefont{Hall, Pinner, and
  Ruderman}}]{Hall:2011aa}
\bibinfo{author}{\bibfnamefont{L.~J.} \bibnamefont{Hall}},
  \bibinfo{author}{\bibfnamefont{D.}~\bibnamefont{Pinner}}, \bibnamefont{and}
  \bibinfo{author}{\bibfnamefont{J.~T.} \bibnamefont{Ruderman}}
  (\bibinfo{year}{2011}), \eprint{1112.2703}.

\bibitem[{\citenamefont{Arbey et~al.}(2011{\natexlab{b}})\citenamefont{Arbey,
  Battaglia, and Mahmoudi}}]{Arbey:2011aa}
\bibinfo{author}{\bibfnamefont{A.}~\bibnamefont{Arbey}},
  \bibinfo{author}{\bibfnamefont{M.}~\bibnamefont{Battaglia}},
  \bibnamefont{and} \bibinfo{author}{\bibfnamefont{F.}~\bibnamefont{Mahmoudi}}
  (\bibinfo{year}{2011}{\natexlab{b}}), \eprint{1112.3032}.

\bibitem[{\citenamefont{Albornoz~Vasquez
  et~al.}(2011)\citenamefont{Albornoz~Vasquez, Belanger, Godbole, and
  Pukhov}}]{AlbornozVasquez:2011aa}
\bibinfo{author}{\bibfnamefont{D.}~\bibnamefont{Albornoz~Vasquez}},
  \bibinfo{author}{\bibfnamefont{G.}~\bibnamefont{Belanger}},
  \bibinfo{author}{\bibfnamefont{R.}~\bibnamefont{Godbole}}, \bibnamefont{and}
  \bibinfo{author}{\bibfnamefont{A.}~\bibnamefont{Pukhov}}
  (\bibinfo{year}{2011}), \eprint{1112.2200}.

\bibitem[{\citenamefont{B\'elanger et~al.}(2001)\citenamefont{B\'elanger,
  Boudjema, Cottrant, Godbole, and Semenov}}]{Belanger:2001am}
\bibinfo{author}{\bibfnamefont{G.}~\bibnamefont{B\'elanger}},
  \bibinfo{author}{\bibfnamefont{F.}~\bibnamefont{Boudjema}},
  \bibinfo{author}{\bibfnamefont{A.}~\bibnamefont{Cottrant}},
  \bibinfo{author}{\bibfnamefont{R.}~\bibnamefont{Godbole}}, \bibnamefont{and}
  \bibinfo{author}{\bibfnamefont{A.}~\bibnamefont{Semenov}},
  \bibinfo{journal}{Phys.Lett.} \textbf{\bibinfo{volume}{B519}},
  \bibinfo{pages}{93} (\bibinfo{year}{2001}), \eprint{hep-ph/0106275}.

\bibitem[{\citenamefont{Englert
  et~al.}(2011{\natexlab{a}})\citenamefont{Englert, Jaeckel, Re, and
  Spannowsky}}]{Englert:2011us}
\bibinfo{author}{\bibfnamefont{C.}~\bibnamefont{Englert}},
  \bibinfo{author}{\bibfnamefont{J.}~\bibnamefont{Jaeckel}},
  \bibinfo{author}{\bibfnamefont{E.}~\bibnamefont{Re}}, \bibnamefont{and}
  \bibinfo{author}{\bibfnamefont{M.}~\bibnamefont{Spannowsky}}
  (\bibinfo{year}{2011}{\natexlab{a}}), \eprint{1111.1719}.

\bibitem[{\citenamefont{Abel et~al.}(1993)\citenamefont{Abel, Sarkar, and
  Whittingham}}]{Abel:1992ts}
\bibinfo{author}{\bibfnamefont{S.}~\bibnamefont{Abel}},
  \bibinfo{author}{\bibfnamefont{S.}~\bibnamefont{Sarkar}}, \bibnamefont{and}
  \bibinfo{author}{\bibfnamefont{I.}~\bibnamefont{Whittingham}},
  \bibinfo{journal}{Nucl.Phys.} \textbf{\bibinfo{volume}{B392}},
  \bibinfo{pages}{83} (\bibinfo{year}{1993}), \eprint{hep-ph/9209292}.

\bibitem[{\citenamefont{Ellwanger et~al.}(2010)\citenamefont{Ellwanger,
  Hugonie, and Teixeira}}]{Ellwanger:2009dp}
\bibinfo{author}{\bibfnamefont{U.}~\bibnamefont{Ellwanger}},
  \bibinfo{author}{\bibfnamefont{C.}~\bibnamefont{Hugonie}}, \bibnamefont{and}
  \bibinfo{author}{\bibfnamefont{A.~M.} \bibnamefont{Teixeira}},
  \bibinfo{journal}{Phys.Rept.} \textbf{\bibinfo{volume}{496}},
  \bibinfo{pages}{1} (\bibinfo{year}{2010}), \eprint{0910.1785}.

\bibitem[{\citenamefont{Ellwanger et~al.}(2011)\citenamefont{Ellwanger,
  Espitalier-Noel, and Hugonie}}]{Ellwanger:2011mu}
\bibinfo{author}{\bibfnamefont{U.}~\bibnamefont{Ellwanger}},
  \bibinfo{author}{\bibfnamefont{G.}~\bibnamefont{Espitalier-Noel}},
  \bibnamefont{and} \bibinfo{author}{\bibfnamefont{C.}~\bibnamefont{Hugonie}},
  \bibinfo{journal}{JHEP} \textbf{\bibinfo{volume}{1109}}, \bibinfo{pages}{105}
  (\bibinfo{year}{2011}), \eprint{1107.2472}.

\bibitem[{\citenamefont{Ellwanger}(2011{\natexlab{a}})}]{Ellwanger:2010nf}
\bibinfo{author}{\bibfnamefont{U.}~\bibnamefont{Ellwanger}},
  \bibinfo{journal}{Phys.Lett.} \textbf{\bibinfo{volume}{B698}},
  \bibinfo{pages}{293} (\bibinfo{year}{2011}{\natexlab{a}}),
  \eprint{1012.1201}.

\bibitem[{\citenamefont{Ellwanger}(2011{\natexlab{b}})}]{Ellwanger:2011aa}
\bibinfo{author}{\bibfnamefont{U.}~\bibnamefont{Ellwanger}}
  (\bibinfo{year}{2011}{\natexlab{b}}), \eprint{1112.3548}.

\bibitem[{\citenamefont{Kang et~al.}(2012)\citenamefont{Kang, Li, and
  Li}}]{Kang:2012tn}
\bibinfo{author}{\bibfnamefont{Z.}~\bibnamefont{Kang}},
  \bibinfo{author}{\bibfnamefont{J.}~\bibnamefont{Li}}, \bibnamefont{and}
  \bibinfo{author}{\bibfnamefont{T.}~\bibnamefont{Li}} (\bibinfo{year}{2012}),
  \eprint{1201.5305}.

\bibitem[{\citenamefont{Aprile et~al.}(2011)}]{Aprile:2011hi}
\bibinfo{author}{\bibfnamefont{E.}~\bibnamefont{Aprile}} \bibnamefont{et~al.}
  (\bibinfo{collaboration}{XENON100}), \bibinfo{journal}{Phys.Rev.Lett.}
  \textbf{\bibinfo{volume}{107}}, \bibinfo{pages}{131302}
  (\bibinfo{year}{2011}), \eprint{1104.2549}.

\bibitem[{\citenamefont{Abdo et~al.}(2010)\citenamefont{Abdo, Ackermann,
  Ajello, Atwood, Baldini et~al.}}]{Abdo:2010ex}
\bibinfo{author}{\bibfnamefont{A.}~\bibnamefont{Abdo}},
  \bibinfo{author}{\bibfnamefont{M.}~\bibnamefont{Ackermann}},
  \bibinfo{author}{\bibfnamefont{M.}~\bibnamefont{Ajello}},
  \bibinfo{author}{\bibfnamefont{W.}~\bibnamefont{Atwood}},
  \bibinfo{author}{\bibfnamefont{L.}~\bibnamefont{Baldini}},
  \bibnamefont{et~al.}, \bibinfo{journal}{Astrophys.J.}
  \textbf{\bibinfo{volume}{712}}, \bibinfo{pages}{147} (\bibinfo{year}{2010}),
  \eprint{1001.4531}.

\bibitem[{\citenamefont{Albornoz~V\'asquez
  et~al.}(2011)\citenamefont{Albornoz~V\'asquez, B\'elanger, and
  Boehm}}]{AlbornozVasquez:2011js}
\bibinfo{author}{\bibfnamefont{D.}~\bibnamefont{Albornoz~V\'asquez}},
  \bibinfo{author}{\bibfnamefont{G.}~\bibnamefont{B\'elanger}},
  \bibnamefont{and} \bibinfo{author}{\bibfnamefont{C.}~\bibnamefont{Boehm}},
  \bibinfo{journal}{Phys.Rev.} \textbf{\bibinfo{volume}{D84}},
  \bibinfo{pages}{095008} (\bibinfo{year}{2011}), \eprint{1107.1614}.

\bibitem[{\citenamefont{Albornoz~V\'asquez
  et~al.}(2010)\citenamefont{Albornoz~V\'asquez, B\'elanger, Boehm, Pukhov, and
  Silk}}]{Vasquez:2010ru}
\bibinfo{author}{\bibfnamefont{D.}~\bibnamefont{Albornoz~V\'asquez}},
  \bibinfo{author}{\bibfnamefont{G.}~\bibnamefont{B\'elanger}},
  \bibinfo{author}{\bibfnamefont{C.}~\bibnamefont{Boehm}},
  \bibinfo{author}{\bibfnamefont{A.}~\bibnamefont{Pukhov}}, \bibnamefont{and}
  \bibinfo{author}{\bibfnamefont{J.}~\bibnamefont{Silk}},
  \bibinfo{journal}{Phys.Rev.} \textbf{\bibinfo{volume}{D82}},
  \bibinfo{pages}{115027} (\bibinfo{year}{2010}), \eprint{1009.4380}.

\bibitem[{\citenamefont{Vasquez et~al.}(2012)\citenamefont{Vasquez, Belanger,
  Billard, and Mayet}}]{Vasquez:2012px}
\bibinfo{author}{\bibfnamefont{D.}~\bibnamefont{Vasquez}},
  \bibinfo{author}{\bibfnamefont{G.}~\bibnamefont{Belanger}},
  \bibinfo{author}{\bibfnamefont{J.}~\bibnamefont{Billard}}, \bibnamefont{and}
  \bibinfo{author}{\bibfnamefont{F.}~\bibnamefont{Mayet}}
  \bibinfo{journal}{Phys.Rev.} \textbf{\bibinfo{volume}{D85}},
  \bibinfo{pages}{055023} (\bibinfo{year}{2012}), \eprint{1201.6150}.

\bibitem[{\citenamefont{Ellwanger and Hugonie}(2006)}]{Ellwanger:2005dv}
\bibinfo{author}{\bibfnamefont{U.}~\bibnamefont{Ellwanger}} \bibnamefont{and}
  \bibinfo{author}{\bibfnamefont{C.}~\bibnamefont{Hugonie}},
  \bibinfo{journal}{Comput.Phys.Commun.} \textbf{\bibinfo{volume}{175}},
  \bibinfo{pages}{290} (\bibinfo{year}{2006}), \eprint{hep-ph/0508022}.

\bibitem[{\citenamefont{Strigari et~al.}(2007)\citenamefont{Strigari,
  Koushiappas, Bullock, and Kaplinghat}}]{Strigari:2006rd}
\bibinfo{author}{\bibfnamefont{L.~E.} \bibnamefont{Strigari}},
  \bibinfo{author}{\bibfnamefont{S.~M.} \bibnamefont{Koushiappas}},
  \bibinfo{author}{\bibfnamefont{J.~S.} \bibnamefont{Bullock}},
  \bibnamefont{and}
  \bibinfo{author}{\bibfnamefont{M.}~\bibnamefont{Kaplinghat}},
  \bibinfo{journal}{Phys.Rev.} \textbf{\bibinfo{volume}{D75}},
  \bibinfo{pages}{083526} (\bibinfo{year}{2007}), \eprint{astro-ph/0611925}.

\bibitem[{\citenamefont{Boehm et~al.}(2004)\citenamefont{Boehm, Ensslin, and
  Silk}}]{Boehm:2002yz}
\bibinfo{author}{\bibfnamefont{C.}~\bibnamefont{Boehm}},
  \bibinfo{author}{\bibfnamefont{T.~A.} \bibnamefont{Ensslin}},
  \bibnamefont{and} \bibinfo{author}{\bibfnamefont{J.}~\bibnamefont{Silk}},
  \bibinfo{journal}{J. Phys.} \textbf{\bibinfo{volume}{G30}},
  \bibinfo{pages}{279} (\bibinfo{year}{2004}), \eprint{astro-ph/0208458}.

\bibitem[{\citenamefont{Boehm et~al.}(2010)\citenamefont{Boehm, Silk, and
  Ensslin}}]{Boehm:2010kg}
\bibinfo{author}{\bibfnamefont{C.}~\bibnamefont{Boehm}},
  \bibinfo{author}{\bibfnamefont{J.}~\bibnamefont{Silk}}, \bibnamefont{and}
  \bibinfo{author}{\bibfnamefont{T.}~\bibnamefont{Ensslin}}
  (\bibinfo{year}{2010}), \eprint{1008.5175}.

\bibitem[{\citenamefont{Seidel}()}]{Cresst}
\bibinfo{author}{\bibfnamefont{A.}~\bibnamefont{Seidel}},
  \emph{\bibinfo{title}{{Higgs physics at the LHC and ILC}}},
  \bibinfo{note}{{plenary talk at Identification of Dark Matter (IDM10), 26
  July 2010, Montpellier, France.}}

\bibitem[{\citenamefont{Aalseth et~al.}(2010)}]{Aalseth:2010vx}
\bibinfo{author}{\bibfnamefont{C.~E.} \bibnamefont{Aalseth}}
  \bibnamefont{et~al.} (\bibinfo{collaboration}{CoGeNT})
  (\bibinfo{year}{2010}), \eprint{1002.4703}.

\bibitem[{\citenamefont{Bernabei et~al.}(2010)}]{Bernabei:2010mq}
\bibinfo{author}{\bibfnamefont{R.}~\bibnamefont{Bernabei}}
  \bibnamefont{et~al.}, \bibinfo{journal}{Eur. Phys. J.}
  \textbf{\bibinfo{volume}{C67}}, \bibinfo{pages}{39} (\bibinfo{year}{2010}),
  \eprint{1002.1028}.
  
  \bibitem{Aaij:2012ac}
  R.~Aaij {\it et al.}  [LHCb Collaboration],
  Phys.\ Rev.\ Lett.\  {\bf 108} (2012) 231801,
  \eprint{1203.4493}.
 

\bibitem[{\citenamefont{B\'elanger et~al.}(2011)\citenamefont{B\'elanger,
  Boudjema, Brun, Pukhov, Rosier-Lees et~al.}}]{Belanger:2010gh}
\bibinfo{author}{\bibfnamefont{G.}~\bibnamefont{B\'elanger}},
  \bibinfo{author}{\bibfnamefont{F.}~\bibnamefont{Boudjema}},
  \bibinfo{author}{\bibfnamefont{P.}~\bibnamefont{Brun}},
  \bibinfo{author}{\bibfnamefont{A.}~\bibnamefont{Pukhov}},
  \bibinfo{author}{\bibfnamefont{S.}~\bibnamefont{Rosier-Lees}},
  \bibnamefont{et~al.}, \bibinfo{journal}{Comput.Phys.Commun.}
  \textbf{\bibinfo{volume}{182}}, \bibinfo{pages}{842} (\bibinfo{year}{2011}),
  \eprint{1004.1092}.

\bibitem[{\citenamefont{Komatsu et~al.}(2009)}]{Komatsu:2008hk}
\bibinfo{author}{\bibfnamefont{E.}~\bibnamefont{Komatsu}} \bibnamefont{et~al.}
  (\bibinfo{collaboration}{WMAP}), \bibinfo{journal}{Astrophys. J. Suppl.}
  \textbf{\bibinfo{volume}{180}}, \bibinfo{pages}{330} (\bibinfo{year}{2009}),
  \eprint{0803.0547}.

\bibitem[{\citenamefont{Bechtle et~al.}(2010)\citenamefont{Bechtle, Brein,
  Heinemeyer, Weiglein, and Williams}}]{Bechtle:2008jh}
\bibinfo{author}{\bibfnamefont{P.}~\bibnamefont{Bechtle}},
  \bibinfo{author}{\bibfnamefont{O.}~\bibnamefont{Brein}},
  \bibinfo{author}{\bibfnamefont{S.}~\bibnamefont{Heinemeyer}},
  \bibinfo{author}{\bibfnamefont{G.}~\bibnamefont{Weiglein}}, \bibnamefont{and}
  \bibinfo{author}{\bibfnamefont{K.~E.} \bibnamefont{Williams}},
  \bibinfo{journal}{Comput.Phys.Commun.} \textbf{\bibinfo{volume}{181}},
  \bibinfo{pages}{138} (\bibinfo{year}{2010}), \eprint{0811.4169}.

\bibitem[{\citenamefont{Bechtle et~al.}(2011)\citenamefont{Bechtle, Brein,
  Heinemeyer, Weiglein, and Williams}}]{Bechtle:2011sb}
\bibinfo{author}{\bibfnamefont{P.}~\bibnamefont{Bechtle}},
  \bibinfo{author}{\bibfnamefont{O.}~\bibnamefont{Brein}},
  \bibinfo{author}{\bibfnamefont{S.}~\bibnamefont{Heinemeyer}},
  \bibinfo{author}{\bibfnamefont{G.}~\bibnamefont{Weiglein}}, \bibnamefont{and}
  \bibinfo{author}{\bibfnamefont{K.~E.} \bibnamefont{Williams}},
  \bibinfo{journal}{Comput.Phys.Commun.} \textbf{\bibinfo{volume}{182}},
  \bibinfo{pages}{2605} (\bibinfo{year}{2011}), \eprint{1102.1898}.

\bibitem[{\citenamefont{{The ATLAS Collaboration}}(2011)}]{atlas:2011}
\bibinfo{author}{\bibnamefont{{The ATLAS Collaboration}}}
  (\bibinfo{year}{2011}), \bibinfo{note}{aTLAS-CONF-2011-163}.

\bibitem[{\citenamefont{{The CMS Collaboration}}(2011)}]{cms:2011}
\bibinfo{author}{\bibnamefont{{The CMS Collaboration}}} (\bibinfo{year}{2011}),
  \bibinfo{note}{cMS PAS HIG-11-032}.

\bibitem[{\citenamefont{Aad et~al.}(2011)}]{Aad:2011ib}
\bibinfo{author}{\bibfnamefont{G.}~\bibnamefont{Aad}} \bibnamefont{et~al.}
  (\bibinfo{collaboration}{ATLAS Collaboration}) (\bibinfo{year}{2011}),
  \eprint{1109.6572}.

\bibitem[{\citenamefont{Bahr et~al.}(2008)\citenamefont{Bahr, Gieseke, Gigg,
  Grellscheid, Hamilton et~al.}}]{Bahr:2008pv}
\bibinfo{author}{\bibfnamefont{M.}~\bibnamefont{Bahr}},
  \bibinfo{author}{\bibfnamefont{S.}~\bibnamefont{Gieseke}},
  \bibinfo{author}{\bibfnamefont{M.}~\bibnamefont{Gigg}},
  \bibinfo{author}{\bibfnamefont{D.}~\bibnamefont{Grellscheid}},
  \bibinfo{author}{\bibfnamefont{K.}~\bibnamefont{Hamilton}},
  \bibnamefont{et~al.}, \bibinfo{journal}{Eur.Phys.J.}
  \textbf{\bibinfo{volume}{C58}}, \bibinfo{pages}{639} (\bibinfo{year}{2008}),
  \eprint{0803.0883}.

\bibitem[{\citenamefont{Gieseke et~al.}(2011)\citenamefont{Gieseke,
  Grellscheid, Hamilton, Papaefstathiou, Platzer et~al.}}]{Gieseke:2011na}
\bibinfo{author}{\bibfnamefont{S.}~\bibnamefont{Gieseke}},
  \bibinfo{author}{\bibfnamefont{D.}~\bibnamefont{Grellscheid}},
  \bibinfo{author}{\bibfnamefont{K.}~\bibnamefont{Hamilton}},
  \bibinfo{author}{\bibfnamefont{A.}~\bibnamefont{Papaefstathiou}},
  \bibinfo{author}{\bibfnamefont{S.}~\bibnamefont{Platzer}},
  \bibnamefont{et~al.} (\bibinfo{year}{2011}), \eprint{1102.1672}.

\bibitem[{\citenamefont{Buckley et~al.}(2010)\citenamefont{Buckley,
  Butterworth, Lonnblad, Hoeth, Monk et~al.}}]{Buckley:2010ar}
\bibinfo{author}{\bibfnamefont{A.}~\bibnamefont{Buckley}},
  \bibinfo{author}{\bibfnamefont{J.}~\bibnamefont{Butterworth}},
  \bibinfo{author}{\bibfnamefont{L.}~\bibnamefont{Lonnblad}},
  \bibinfo{author}{\bibfnamefont{H.}~\bibnamefont{Hoeth}},
  \bibinfo{author}{\bibfnamefont{J.}~\bibnamefont{Monk}}, \bibnamefont{et~al.}
  (\bibinfo{year}{2010}), \eprint{1003.0694}.

\bibitem[{\citenamefont{Grellscheid et~al.}(2011)\citenamefont{Grellscheid,
  Jaeckel, Khoze, Richardson, and Wymant}}]{Grellscheid:2011ij}
\bibinfo{author}{\bibfnamefont{D.}~\bibnamefont{Grellscheid}},
  \bibinfo{author}{\bibfnamefont{J.}~\bibnamefont{Jaeckel}},
  \bibinfo{author}{\bibfnamefont{V.~V.} \bibnamefont{Khoze}},
  \bibinfo{author}{\bibfnamefont{P.}~\bibnamefont{Richardson}},
  \bibnamefont{and} \bibinfo{author}{\bibfnamefont{C.}~\bibnamefont{Wymant}}
  (\bibinfo{year}{2011}), \eprint{1111.3365}.

\bibitem[{\citenamefont{Das et~al.}(2012)\citenamefont{Das, Ellwanger, and
  Teixeira}}]{Das:2012rr}
\bibinfo{author}{\bibfnamefont{D.}~\bibnamefont{Das}},
  \bibinfo{author}{\bibfnamefont{U.}~\bibnamefont{Ellwanger}},
  \bibnamefont{and} \bibinfo{author}{\bibfnamefont{A.~M.}
  \bibnamefont{Teixeira}} (\bibinfo{year}{2012}), \eprint{1202.5244}.

\bibitem[{\citenamefont{Cao et~al.}(2011{\natexlab{b}})\citenamefont{Cao,
  Hikasa, Wang, Yang, Hikasa et~al.}}]{Cao:2011re}
\bibinfo{author}{\bibfnamefont{J.-J.} \bibnamefont{Cao}},
  \bibinfo{author}{\bibfnamefont{K.-i.} \bibnamefont{Hikasa}},
  \bibinfo{author}{\bibfnamefont{W.}~\bibnamefont{Wang}},
  \bibinfo{author}{\bibfnamefont{J.~M.} \bibnamefont{Yang}},
  \bibinfo{author}{\bibfnamefont{K.-i.} \bibnamefont{Hikasa}},
  \bibnamefont{et~al.}, \bibinfo{journal}{Phys.Lett.}
  \textbf{\bibinfo{volume}{B703}}, \bibinfo{pages}{292}
  (\bibinfo{year}{2011}{\natexlab{b}}), \eprint{1104.1754}.

\bibitem[{\citenamefont{Englert
  et~al.}(2011{\natexlab{b}})\citenamefont{Englert, Roy, and
  Spannowsky}}]{Englert:2011iz}
\bibinfo{author}{\bibfnamefont{C.}~\bibnamefont{Englert}},
  \bibinfo{author}{\bibfnamefont{T.~S.} \bibnamefont{Roy}}, \bibnamefont{and}
  \bibinfo{author}{\bibfnamefont{M.}~\bibnamefont{Spannowsky}},
  \bibinfo{journal}{Phys.Rev.} \textbf{\bibinfo{volume}{D84}},
  \bibinfo{pages}{075026} (\bibinfo{year}{2011}{\natexlab{b}}),
  \eprint{1106.4545}.

\bibitem[{\citenamefont{Lisanti and Wacker}(2009)}]{Lisanti:2009uy}
\bibinfo{author}{\bibfnamefont{M.}~\bibnamefont{Lisanti}} \bibnamefont{and}
  \bibinfo{author}{\bibfnamefont{J.~G.} \bibnamefont{Wacker}},
  \bibinfo{journal}{Phys.Rev.} \textbf{\bibinfo{volume}{D79}},
  \bibinfo{pages}{115006} (\bibinfo{year}{2009}), \eprint{0903.1377}.

\bibitem[{\citenamefont{Falkowski et~al.}(2011)\citenamefont{Falkowski, Krohn,
  Wang, Shelton, and Thalapillil}}]{Falkowski:2010hi}
\bibinfo{author}{\bibfnamefont{A.}~\bibnamefont{Falkowski}},
  \bibinfo{author}{\bibfnamefont{D.}~\bibnamefont{Krohn}},
  \bibinfo{author}{\bibfnamefont{L.-T.} \bibnamefont{Wang}},
  \bibinfo{author}{\bibfnamefont{J.}~\bibnamefont{Shelton}}, \bibnamefont{and}
  \bibinfo{author}{\bibfnamefont{A.}~\bibnamefont{Thalapillil}},
  \bibinfo{journal}{Phys.Rev.} \textbf{\bibinfo{volume}{D84}},
  \bibinfo{pages}{074022} (\bibinfo{year}{2011}), \eprint{1006.1650}.

\bibitem[{\citenamefont{Chen et~al.}(2010)\citenamefont{Chen, Nojiri, and
  Sreethawong}}]{Chen:2010wk}
\bibinfo{author}{\bibfnamefont{C.-R.} \bibnamefont{Chen}},
  \bibinfo{author}{\bibfnamefont{M.~M.} \bibnamefont{Nojiri}},
  \bibnamefont{and}
  \bibinfo{author}{\bibfnamefont{W.}~\bibnamefont{Sreethawong}},
  \bibinfo{journal}{JHEP} \textbf{\bibinfo{volume}{1011}}, \bibinfo{pages}{012}
  (\bibinfo{year}{2010}), \eprint{1006.1151}.

\bibitem[{\citenamefont{Cao et~al.}(2012)\citenamefont{Cao, Heng, Yang, Zhang,
  and Zhu}}]{Cao:2012fz}
\bibinfo{author}{\bibfnamefont{J.}~\bibnamefont{Cao}},
  \bibinfo{author}{\bibfnamefont{Z.}~\bibnamefont{Heng}},
  \bibinfo{author}{\bibfnamefont{J.~M.} \bibnamefont{Yang}},
  \bibinfo{author}{\bibfnamefont{Y.}~\bibnamefont{Zhang}}, \bibnamefont{and}
  \bibinfo{author}{\bibfnamefont{J.}~\bibnamefont{Zhu}} (\bibinfo{year}{2012}),
  \eprint{1202.5821}.

\bibitem[{\citenamefont{Belanger et~al.}(2009)\citenamefont{Belanger, Hugonie,
  and Pukhov}}]{Belanger:2008nt}
\bibinfo{author}{\bibfnamefont{G.}~\bibnamefont{Belanger}},
  \bibinfo{author}{\bibfnamefont{C.}~\bibnamefont{Hugonie}}, \bibnamefont{and}
  \bibinfo{author}{\bibfnamefont{A.}~\bibnamefont{Pukhov}},
  \bibinfo{journal}{JCAP} \textbf{\bibinfo{volume}{0901}}, \bibinfo{pages}{023}
  (\bibinfo{year}{2009}), \eprint{0811.3224}.

\bibitem[{\citenamefont{{Hernando Morata, J.}}(2012)}]{LHCb}
\bibinfo{author}{\bibnamefont{{Hernando Morata, J.}}},
  \emph{\bibinfo{title}{{Rare decays in LHCb, talk presented at 47th Rencontres
  de Moriond, LaThuile, Italy, March 2012.}}} (\bibinfo{year}{2012}),
  \bibinfo{note}{lHCb-TALK-2012-028}.

\bibitem[{\citenamefont{Kraml et~al.}(2009)\citenamefont{Kraml, Raklev, and
  White}}]{Kraml:2008zr}
\bibinfo{author}{\bibfnamefont{S.}~\bibnamefont{Kraml}},
  \bibinfo{author}{\bibfnamefont{A.}~\bibnamefont{Raklev}}, \bibnamefont{and}
  \bibinfo{author}{\bibfnamefont{M.}~\bibnamefont{White}},
  \bibinfo{journal}{Phys.Lett.} \textbf{\bibinfo{volume}{B672}},
  \bibinfo{pages}{361} (\bibinfo{year}{2009}), \eprint{0811.0011}.

\end{thebibliography}

\end{document}